\documentclass[aps,twocolumn,amsmath,amssymb,preprintnumbers]{revtex4}
\usepackage{amsmath} \usepackage{amsfonts} \usepackage{amssymb}
\usepackage{bbm}
\usepackage{epsfig}
\usepackage{graphics}
\usepackage{graphicx}
\newcommand{\be}{\begin{equation}}
\newcommand{\ee}{\end{equation}}
\newcommand{\ba}{\begin{eqnarray}}
\newcommand{\ea}{\end{eqnarray}}
\newcommand{\nn}{\nonumber}
\newcommand{\kr}{\rangle}
\newcommand{\kl}{\langle}
\newcommand{\A}{ A^{(k)}}
\newcommand{\hA}{ \hat A^{(k)}}
\newcommand{\tr}{\textup{tr}}
\newcommand{\ab}{_{\alpha\beta}}

\newcommand{\C}{_{cl}}

\begin{document}

\title[ ]{Quantum particles from classical statistics}

\author{C. Wetterich}
\affiliation{Institut  f\"ur Theoretische Physik\\
Universit\"at Heidelberg\\
Philosophenweg 16, D-69120 Heidelberg}

\begin{abstract}
Quantum particles and classical particles are described in a common setting of classical statistical physics. The property of a particle being ``classical'' or ``quantum'' ceases to be a basic conceptual difference. The dynamics differs, however, between quantum and classical particles. We describe position, motion and correlations of a quantum particle in terms of observables in a classical statistical ensemble. On the other side, we also construct explicitly the quantum formalism with wave function and Hamiltonian for classical particles. For a suitable time evolution of the classical probabilities and a suitable choice of observables all features of a quantum particle in a potential can be derived from classical statistics, including interference and tunneling. Besides conceptual advances, the treatment of classical and quantum particles in a common formalism could lead to interesting cross-fertilization between classical statistics and quantum physics.
\end{abstract}

\maketitle

\section{Introduction}
\label{Intro}
Ever since the appearance of quantum physics it was widely believed that the basic concepts of a classical particle and a quantum particle are fundamentally different and mutually exclusive. In this paper we argue that this is not the case. We describe a quantum particle in a setting of classical statistics with infinitely many degrees of freedom. On the other hand, we describe classical particles, with simultaneously sharp position and momentum, within the formalism of quantum mechanics. Both quantum and classical particles can therefore be described within the same conceptual setting. Their difference resides in the particular dynamics, as expressed by different Hamiltonians which are based on different sets of observables.

Perhaps the most striking evidence that classical particles and quantum particles can be described within the same setting consists in the possibility of ``zwitters'' \cite{3A} - particles with properties interpolating between the quantum particle and the classical particle as a function of some continuous parameter $\gamma$. This parameter parameterizes different possible laws for the time evolution of the probability density in phase space. While for $\gamma=0$ a quantum particle shows interference in a double slit experiment, the classical particle for $\gamma=\pi/2$ passes through only one of the slits without interference effects. Precision measurements can test the ``quantumness'' in a quantitative way by putting bounds on $\gamma$. 

Our implementation of a quantum particle within classical statistics is not a deterministic local hidden variable theory but rather assumes that the fundamental description of the real world is intrinsically probabilistic. We generalize the description of quantum systems with a finite number of states $M$ in terms of classical statistical ensembles \cite{CWAA,CW2,CWE}. The continuous density matrix and wave function of a quantum particle is obtained by taking the limit $M\to\infty$. In the present paper we describe all observables of the quantum particle as standard classical observables that take fixed values for every state of a large classical statistical ensemble. Such an ensemble is described by a classical probability distribution. In two forthcoming papers we will show that a much simpler description in terms of a probability density in usual phase space is possible if one admits the use of ``statistical observables''. Statistical observables are computable in a non-linear way for a given probability density but cannot be associated to quantities taking a fixed value in every state. They are conceptually similar to entropy in thermodynamics. In this setting we can understand the quantum particle as a ``coarse graining'' of a probability distribution of a classical particle. The present work provides the conceptual foundation for these further developments.

The classical statistical foundation of quantum physics is based on several basic concepts  \cite{CWAA,CW2,CWE}: 

(i) The quantum system is described as an isolated subsystem of a classical ensemble. This ensemble involves infinitely many possible states for the particle and its environment. In a sense, the quantum particle is considered as an ``excitation'' of the ``vacuum'' (environment). 

(ii) The subsystem can be characterized by a restricted set of probabilistic observables, for which the spectrum of measurement values and the probabilities to find a given value can be computed from the information which specifies the state of the quantum system. 

(iii) In turn, the state of the quantum system can be characterized by the expectation values of a subset of these ``quantum observables''. 

(iv) Conditional correlations, which are computable from the information contained in the quantum state the particle, are used for predictions of the outcome of measurements of pairs of observables $A$ and $B$. The conditional correlation or ``measurement correlation'' $\kl AB\kr_m$ differs from the classical or pointwise correlation $\kl A\cdot B\kr$. 

(v) In general, the joint probability of finding the value $a$ for $A$ and $b$ for $B$ cannot be computed from the quantum state. Even if joint probabilities exist, they typically involve information which characterizes the environment, in addition to the information characterizing the state of the quantum system. Concentrating on the quantum subsystem, we deal with ``incomplete statistics'' \cite{3}, where joint probabilities are not available or not used for $\kl AB\kr_m$ for all pairs of observables. We will find that a classical particle corresponds to the special case of complete statistics for position and momentum. Only for classical particles enough information is available such that joint probabilities can be used for $\kl AB\kr_m$ for all pairs of particle observables.

(vi) Many different classical observables of the statistical ensemble, which describe the quantum system and its environment, are mapped to the same probabilistic quantum observable. Their difference characterizes different properties of the environment, while they give identical results for measurements concerning only the subsystem properties. The particle observables can be characterized by equivalence classes for the observables of the total classical statistical ensemble. The measurement correlation $\kl AB\kr_m$ is a property of the equivalence classes, while the joint probabilities and the classical correlation $\kl A\cdot B\kr$ depend, in general, on the particular representatives of the equivalence classes or, in other words, on the environment. 

(vii) A time evolution which conserves certain statistical quantities, as the purity and the copurity \cite{CWE} of the subsystem, leads to the unitary time evolution which is characteristic for quantum mechanics. A more general time evolution can also describe decoherence \cite{DC} or syncoherence \cite{CW2}. 

In ref. \cite{CWAA,CW2,CWE,3A} the basic conceptual settings have been described in detail for statistical ensembles which correspond to two-state and four-state quantum mechanics. It was shown how the quantum mechanical formalism with non-commuting operators arises form a description of a subsystem which obeys the basic concepts (i)-(vii). We have discussed explicitly quantum mechanically entangled states \cite{CW1}, \cite{CW2}, \cite{CWE} and shown that the measurement correlation $\kl AB\kr_m$ is equivalent to the usual quantum correlation and violates Bell's inequalities \cite{Bell}. In this context the property of incomplete statistics \cite{3}, where the measurement correlation is not based on joint probabilities, is crucial \cite{CW2}, \cite{CWAA}. Complete statistical systems, for which the measurement correlation employs the joint probabilities, have to obey Bell's inequalities \cite{BS}. Usual local hidden variable theories are assumed to be described by complete statistical systems. In contrast, for a fundamental probabilistic setting \cite{GenStat} the assumption of a complete statistical system describing the subsystem can lead to severe problems \cite{CWAA}. This strongly suggests that incomplete statistics \cite{3} is appropriate. Only in this case the EPR-paradoxon \cite{EPR} can be resolved satisfactorily \cite{CWAA}, \cite{CW2}, and the entanglement, which is the basis for spectacular experiments on teleportation or quantum cryptography \cite{Ze}, or for quantum computing \cite{Zo}, can be explained. The use of probabilistic observables \cite{PO} for the subsystem avoids conflicts with the Kochen-Specker theorem \cite{KS}, as demonstrated explicitly in \cite{CWAA}.

In the present paper we will not repeat the general discussion of concepts and refer to \cite{CWAA}. Our purpose will be the generalization to quantum systems with a continuous family of quantum states, as appropriate for a quantum particle whose wave function depends on a continuous variable as position or momentum. For this purpose we will first characterize the state of the particle-subsystem by a finite number of possibilities for the outcome of yes/no-questions. Such probabilistic systems, where the information concerns only $Q$ bits, correspond to $2^Q$-state quantum systems. We will then take the limit $Q\to\infty$ in order to construct the continuous location and momentum observables for a particle. The start from a finite number of $Q$ bits makes the possibility of incompleteness of the statistical description of the subsystem particularly apparent.

In fact, the difference between a quantum particle and a classical particle precisely reflects the issue of incompleteness or completeness of the statistical description. For a quantum particle the joint probabilities for answers to location and motion questions are not available. This leads to non-commuting operators describing the position and the momentum of a particle and to Heisenberg's uncertainty relation. In contrast, a different set of yes/no-questions, for which joint probabilities for the outcome of location and motion questions can be given, describes a classical particle. Now the position and momentum operators commute and a simultaneous sharp measurement for both types of observables becomes possible. For both the quantum particle and the classical particle the formalism of quantum mechanics can be used. The description of the classical particle employs, however, an unusual set of observables where location and momentum operators commute. 

This paper is organized as follows: In sect. \ref{particlelocation} we discuss the position or location observable as an example of how to obtain observables with a continuous spectrum from the limit $Q\to\infty$. In sect. \ref{quantumandclassical} we then turn to alternative questions about motion and correlation for the particle. At this level the incompleteness characteristic for quantum particles plays a role. In sects. \ref{particlemotion} and \ref{particlecorrelation} we discuss the observables describing the motion of the particle (momentum) and the ``particle-correlation'', which is typical for a quantum mechanical wave packet. In sect. \ref{classicalparticle} we turn to the quantum mechanical description of a classical particle with commuting position and momentum operators. We specify the Hamiltonian which leads to classical trajectories and show that the time evolution of the wave function implies the time evolution of the probability distribution according to the Liouville equation for pointlike particles on classical trajectories. We compare the time evolution of quantum and classical particles in sect. \ref{inapotential}. Sect. \ref{Quantumparticlesfrom} deepens the connection between classical and quantum particles by using probabilities in classical phase space for the description of quantum particles. This issue will be discussed more extensively in two forthcoming papers. Conclusion and discussion follow in sect. \ref{conclusionsand}. 

\section{Particle location}
\label{particlelocation}
Consider a particle trapped in some region of space as, for example, a cavity or an atom trap. A priori we will not assume any property of the particle - it  may be an extended, pointlike of diffuse object, and one may be able to assign to it properties as momentum or angular momentum or not. It will be our task to find out under which conditions certain properties can be used for its characterization. We will assume, however, that the particle and its environment can be described by a classical statistical ensemble. The environment includes the vacuum - which is a complicated entity in the perspective of some more fundamental quantum field theory - and the trap. The classical statistical ensemble will be considered in the limit of infinitely many classical states and the classical ensemble is specified by the probabilities for these states. One may view the particle as some type of excitation of the vacuum in presence of the ``external fields '' (typically electromagnetic and gravitational) which assure the confinement within the trap.

In order to speak about a ``particle'' we consider our object as an ``isolated subsystem'' of the classical statistical ensemble. Isolation is used here in the sense that the expectation values for a finite number of ``basis observables'' are sufficient for the characterization of the state of the particle. All properties of the particle should be expressed in terms of observables whose expectation values and correlations can be computed once the state of the particle is given. The state of the particle requires only a small part of the information contained in the probability distribution characterizing the ensemble of particle and environment - this is the basic reason why we typically have to deal with ``incomplete statistics'' \cite{3}, \cite{CW2}, \cite{CWE}. 

Assume first that the properties of the particle reflect possible answers to a finite number $Q$ of yes/no questions or $Q$ bits. For example, we may use $Q$ bits in order to characterize the location of the particle. For this purpose we ask $Q$ questions of the type: ``is the particle in a given portion of the volume?'', in order to divide the volume of the trap into $M=2^Q$ cells. We enumerate these cells by $\alpha,\alpha=1\dots M$. The knowledge of the state of the subsystem requires then $M$ probabilities $w_\alpha \geq 0,~\sum_\alpha w_\alpha=1$, for the particle being found in the cell $\alpha$. We may define $M$ two-level observables $A^{(\alpha)}$ that take the value $1$ if the particle is in cell $\alpha$ and $-1$ if not, such that $(A^{(\alpha)})^2=1$ for all states of the ensemble. This induces for each cell a particle number $N^{(\alpha)}=\frac12(1+A^{(\alpha)})$ which takes values $1$ or $0$.

We may further associate a cartesian coordinate $x_i(\alpha)$ to each cell - for example $x_i(\alpha)=\int_{V(\alpha)}y_id^3y/\int_{V(\alpha)}d^3y$ with an integration over the volume $V(\alpha)$ of cell $\alpha$. This allows us to define an observable for the location of the particle by
\be\label{1}
X_i=\sum_\alpha x_i(\alpha)N^{(\alpha)},
\ee
with expectation value
\be\label{2}
\langle X_i\rangle =\bar x_i=\sum_\alpha x_i(\alpha)\langle N^{(\alpha)}\rangle=\sum_\alpha x_i(\alpha)w_\alpha.
\ee
Here $\langle N^{(\alpha)}\rangle$ is computed in the classical statistical ensemble of particle and environment. Since $\sum_\alpha N^{(\alpha)}$ equals the unit observable for a particle confined in the volume of the trap, one has indeed $0\leq \langle N^{(\alpha)}\rangle \leq 1,~\sum_\alpha\langle N^{(\alpha)}\rangle =1$, as required for the identification $\langle N^{(\alpha)}\rangle=w_\alpha$. 

We may also define observables for higher moments of the location observable, as the ``dispersion tensor''
\be\label{3}
X^{(2)}_{ij}=\sum_\alpha x_i(\alpha)x_j(\alpha)N^{(\alpha)}-\bar x_i\bar x_j,
\ee
with 
\be\label{4}
\langle  X^{(2)}_{ij}\rangle=\sum_\alpha\big(x_i(\alpha)-\bar x_i\big)
\big(x_j(\alpha)-\bar x_j\big)w_\alpha.
\ee
A ``classical eigenstate'' for the particle numbers consists of an ensemble for which $\langle N^{(\bar\alpha)}\rangle=1,~\langle N^{(\alpha\neq\bar\alpha)}\rangle =0$ for some fixed index $\bar\alpha$. This implies also an eigenstate of $X_i$ with $\bar x_i=x_i(\bar\alpha)$ and vanishing dispersion $\langle X^{(2)}_{ij}\rangle =0$. In contrast, for ``equipartition''  with $\langle N^{(\alpha)}\rangle=w_\alpha=1/M$ and for cells with $\sum_\alpha x_i(\alpha)=0$ one finds $\bar x_i=0$ and $\langle X^{(2)}_{ij}\rangle=\sum_\alpha x_i(\alpha)x_j(\alpha)/M$. For $M\to\infty$ and fixed volume of the trap the location observable $X_i$ becomes continuous. It is a possible candidate for a basis observable. 

We may cast our setting into a formalism familiar from quantum mechanics by associating the particle numbers $N^{(\alpha)}$ with operators $\hat N^{(\alpha)}$, which are defined as diagonal $M\times M$ matrices (no summation over $\alpha$)
\be\label{5}
(\hat N^{(\alpha)})_{\beta\gamma}=\delta_{\beta\alpha}\delta_{\gamma\alpha}.
\ee
We further introduce a hermitean density matrix $\rho$, whose diagonal elements read $\rho_{\alpha\alpha}=w_\alpha$. One infers the quantum mechanical rule for the computation of expectation values in terms of the density matrix
\ba\label{6}
\langle N^{(\alpha)}\rangle&=&\textup{tr}(\rho\hat N^{(\alpha)})=\rho_{\alpha\alpha}=w_\alpha,\nn\\
\langle X_i\rangle&=&\textup{tr}(\rho\hat X_i)~,~
\hat X_i=\sum_\alpha x_i(\alpha)\hat N^{(\alpha)}. 
\ea
At this stage all operators are diagonal and only the diagonal elements of the density matrix matter. There is no difference between ``classical particles'' and ``quantum particles'' up to now. This difference arises only once we ask the question ``What else can we know besides the location of the particle?''

\section{Quantum and classical particle}
\label{quantumandclassical}
In order to investigate the difference between quantum particles and classical particles we simplify our system to be one dimensional, choosing a trap in the form of a ring. The cell coordinates $x(\alpha)$ and the observable $X$ are now periodic in the range $-\pi\leq x(\alpha)\leq \pi$. For the location we can use a hierarchical sequence of bits, the first dividing the ring into its left and right half, the next for subdividing each half into quarters and so on. We will now consider more general types of questions. 

\medskip\noindent
{\bf 1. One-bit-particle}

Let us first investigate a hypothetical ``particle'' that can be described by only one bit, $Q=1,~M=2$. By definition, it must be possible to characterize the state of this particle by only one yes/no question. This description of the subsystem by only one bit is considered as a basic property of the particle, while the probabilities $w_+$ and $w_-$ for finding in an experiment the answers yes or no specify the state of the particle $(w_++w_-=1)$. The ``particle observable'' $A$ is a probabilistic two level observable which can only take the values $\pm 1$ (corresponding to the answers yes or no). In other words, its spectrum consists of two values $\gamma_\alpha=\pm 1$. The probability $w_+$ allows one to compute the expectation values for arbitrary powers of the observable. In particular $\kl A^2\kr=1$ is a sharp value independently of the state characterized by $w_+$. The central issue is: what is the question? 

One possibility is the ``location question'', with associated operators 
\be\label{7}
\hat N^{(1)}=\left(\begin{array}{ll}
1&0\\0&0\end{array}\right)~,~
\hat N^{(2)}=\left(\begin{array}{ll}
0&0\\0&1\end{array}\right)~,~
\hat X=\frac\pi2
\left(\begin{array}{ccl}
1&0\\0&-1\end{array}\right).
\ee
However, other questions may also be asked as we demonstrate next in an example for a ``one-bit-particle''. (This ``Gedankenexperiment'' only illustrates the different possibilities a ``one bit question'' and should not be mistaken as a description of real measurements of a quantum particle.) Suppose that we have placed in the ring two detectors on the right and left sides (say at $x=\pm \pi/2)$. These detectors are assumed to signal $+1$ if the ``particle'' goes through clockwise and $-1$ if it passes anticlockwise. We assume that the particle is not substantially disturbed by the countings of the detectors. Each one of the two detectors $(l)$ (at $x=-\pi/2)$ and $(r)$ (at $x=\pi/2)$ will record a series of $+1$ and $-1$ values. The probabilities to find given series of $\pm 1$ hits reflect the state of particle {\em and} environment, including the measurement apparatus. They contain much more information than contained in one yes/no question. From this information we want to extract the more limited information about the state of the particle using only one yes/no question. 

A measurement analyzes the series of $\pm 1$ values in a certain time interval $\Delta t$, which includes typically a large number of hits in the detectors. One possible two-level-observable for the particle takes the value $+1$ if there are more hits in $(r)$ as compared to $(l)$, and $-1$ in the opposite case. (For simplicity we assume an odd number of hits such that the question has a unique answer.) This ``location question'' can be associated with a ``location observable'' $L$. If there are more hits in the detector $(l)$ than in $(r)$ we define that the particle is in the left half of the ring with $L=-1$. The classical ensemble, which describes the particle and its environment, specifies a probability for each possible series of hits. In turn, this determines the probabilities $w^{(L)}_+$ and $w^{(L)}_-=1-w^{(L)}_+$ to find for the location observable the values $+1$ or $-1$, which describe the state of the particle. If we associate to $L$ the operator $\hat L=\tau_3$ we recognize that $\hat X=(\pi/2)\hat L$ corresponds to the location observable in different units.

However, we may also ask a different question, as the ``motion question'': ``do both detectors $(r)$ and $(l)$ together  show more $+1$ hits than $-1$ hits?''. If yes, the ``motion observable'' $M$ takes the value $+1$ (for clockwise motion), if no, we associate $M=-1$ to anticlockwise motion. Again, the classical ensemble induces probabilities $w^{(M)}_+$ and $w^{(M)}_-=1-w^{(M)}_+$ and therefore an expectation value for the motion observable $\langle M\rangle=w^{(M)}_+-w^{(M)}_-$. The motion observable can be associated with the ``momentum'' of the particle in appropriate units. The motion observable fulfills all criteria for a yes/no question for the one-bit-particle, just as the location observable does.

A third two-level-observable can be associated with the ``correlation question''. We may register a ``jump'' if a hit in $(l)$ is followed by a hit in $(r)$, or if a hit in $(l)$ occurs after a hit in $(r)$. A ``stay'' is a sequence of two hits in the same detector. If a sequence of hits shows more jumps than stays we may assign the value $+1$ to the ``correlation observable'' $C$, with the intuitive notion that for many jumps the particle has some tendency to be both on the left and the right side. In the opposite case of more stays than jumps the observable $C$ assumes the value $-1$, with ``anticorrelation'' associated to the notion that the particle has a tendency to be exclusively either left or right. Again the average of the correlation observable can be expressed in terms of the probabilities $w^{(C)}_\pm$ to find $C=\pm 1$, namely $\langle C\rangle=w^{(C)}_+-w^{(C)}_-$. 

The three two-level-observables $L,M$ and $C$ are unrelated in the sense that the knowledge of the value of one of the observables does not yield any information on the other two observables. If we consider only chains with more hits in $(r)$ than in $(l)$, i.e. $L=+1$, this does not tell us anything on the relative number of $+1$ and $-1$ hits associated to $M$ or on the relative number of jumps and stays associated to $C$. In other words, we assume that the restriction to ensembles with more hits in $(r)$ than $(l)$, i.e. an eigenstate to $L$ with $\langle L\rangle=1~,~w^{(L)}_+=1~,~w^{(L)}_-=0$, does not favor one of the two possible values for $M$ or for $C$. Similarly the eigenstates of $M$ and $C$ should not ``bias'' the outcome of measurements of the two complementary variables. This can be realized for suitable rules for sequences of hits, for example by the ``classical path sequences'' where a jump consists of two hits $+1$ of two hits $-1$ in the two different detectors (but not two hits with opposite sign), while a stay consists of two hits with opposite sign in the same detector (the particle moving forward and backwards through the detector). It also holds for ``random sequences'' where sequences with arbitrary $+1$ and $-1$ hits are allowed.

\medskip\noindent
{\bf 2. Particle subsystem}

On a somewhat more formal level we may associate each series of hits (each possible sequence of $\pm 1$ values) in the two detectors $(r)$ and $(l)$ with a state $\tau$ of the total system of particle and environment. The probabilities $p_\tau$ for individual sequences characterize the classical statistical ensemble. They allow us to compute the probabilities $w^{(L)}_+,w^{(M)}_+$ and $w^{(C)}_+$ for finding the values $+1$ for the two level observables $L,M$ and $C$, or equivalently to compute the expectation values $\kl L\kr,\kl M\kr$ and $\kl C\kr$. We will now assume that the expectation values $\kl L\kr,\kl M\kr,\kl C\kr$ characterize the state of the one-bit-particle, and that this is the {\em only} information available for the subsystem. The joint probability of finding $L=1$ and $M=1$ cannot be extracted from the knowledge of $\kl L\kr, \kl M\kr, \kl C\kr$ - it is not a property of our one-bit particle. The joint probability would be computable from the knowledge of $p_\tau$ for all sequences $\tau$, but this involves properties of the environment. 

In fact, if the joint probabilities for the different combinations $(+,+),(+,-),(-,+)$ and $(-,-)$ for the observables $L$ and $M$ would be available from the state of the particle, we could construct a further observable $LM$ which takes the value $+1$ if $L$ and $M$ have equal sign, and $-1$ for opposite signs. Its expectation value $\kl LM\kr$ would be computable from the particle state. However, $\kl LM\kr$ cannot be computed form $\kl L\kr$ and $\kl M\kr$ alone. Furthermore, the observable $C$ is different from $LM$. We conclude that $\kl LM\kr$ cannot be computed from the knowledge of $\kl L\kr, \kl M\kr$ and $\kl C\kr$, but would require additional ``environmental information'' not available if the particle state is only characterized by $\kl L\kr,\kl M\kr,\kl C\kr$. The statistical system describing the particle alone is incomplete. Furthermore, if $LM$ would be a particle observable, then $L,M$ and $LM$ would be sufficient in order to construct a composite observable with a spectrum of {\em four} different values, corresponding to the $(+,+),(+,-),(-,+)$ and $(-,-)$ combinations for $L$ and $M$. A spectrum with four different values for a particle observable is in contradiction with the assumption that all particle observables use only one bit of information.

Let us consider $L,M$ and $C$ as basis observables for the subsystem. The state of the subsystem is characterized by their expectation values $\langle L\rangle~,~\langle M\rangle$ and $\langle C\rangle$. We may associate to these observables the operators ($2\times 2$ Pauli matrices)
\be\label{8}
\hat L=\tau_3~,~\hat M=-\tau_2~,~\hat C=\tau_1.
\ee
We can further construct a hermitean matrix $\rho$ which characterizes the state of the particle
\be\label{9}
\rho=\frac12 (1+\langle L\rangle\tau_3-\langle M\rangle\tau_2+\langle C\rangle \tau_1).
\ee
It obeys tr$\rho=1$ and $0\leq\rho_{\alpha\alpha}\leq 1$. For all observables $A=(L,M,C)$ the expectation values obey
\be\label{10}
\langle A\rangle =\textup{tr}(\rho\hat A). 
\ee

The third condition for $\rho$ being a density matrix, tr$\rho^2\leq 1$, is obeyed only for $\langle L\rangle^2+\langle M\rangle^2+\langle C\rangle^2\leq 1$. This implies that at most one of the three observables can have a sharp value. For example, $\langle M\rangle=1$ implies $\langle L\rangle=\kl C \kr=0$. This is a typical situation for a quantum particle. For our one bit particle the sum $\kl L\kr^2+\kl M\kr^2+\kl C\kr^2$ describes the purity $P$ of the subsystem. We will assume $P\leq 1$, in accordance with the restriction to one-bit observables. In contrast, a classical particle would have a sharp value for both $L$ and $M$ and therefore $\kl L\kr^2+\kl M\kr^2+\kl C\kr^2\geq 2$. However, simultaneously sharp values of $L$ and $M$ answer simultaneously two yes/no questions (``is $L$ positive?'' and ``is $M$ positive?'') such that this case should be described by a particle characterized by two bits.

\medskip\noindent
{\bf 3. Two-bit-particle}

Let us therefore extend the discussion to a hypothetical particle characterized by two bits, $Q=2~,~M=4$. We generalize the observables $L,M,C$ to fifteen two-level observables $A^{(k)}~,~k=1\dots 15,(A^{(k)})^2=1$. They are considered to be the possible basis observables of the isolated sub-system, such that the state of the particle is described by fifteen real numbers $\rho_k~,~-1\leq\rho_k\leq 1$,
\be\label{8a}
\rho_k=\kl \A\kr.
\ee
We define the purity of the subsystem as
\be\label{9a}
P=\sum_k(\rho_k)^2
\ee
and choose $P\leq 3$, as appropriate for two bits. Indeed, with $Q$ yes/no questions we can fix at most $2^Q-1$ sharp values of independent two level observables, similar to the particle numbers $N^{(\alpha)}$ in section \ref{particlelocation}. (The constraint $\sum_\alpha N^{(\alpha)}=1$ reduces the number of independent observables by one.) 

We introduce the $M^2-1$ normalized $SU(M)$ generators $L_k$ obeying $L^2_k=1$, $L^\dagger_k=L_k$, tr$L_k=0,$ tr$(L_k,L_l)=M\delta_{kl}$ \cite{CW1}. With $|\rho_k|\leq1$ the hermitean matrix 
\be\label{10a}
\rho=\frac{1}{M}(1+\rho_kL_k)
\ee
obeys two of the requirements for a density matrix: tr$\rho=1$, tr$\rho^2=(P+1)/M\leq 1$. The two-level observables $\A$ are represented as hermitean $M\times M$ matrices or ``operators''
\be\label{11}
\hA =L_k~,~\kl\A\kr=\tr(\rho\hA)=\rho_k.
\ee
Their expectation values can be computed according to the rule of quantum mechanics in terms of the density matrix. 

In particular, we consider the diagonal operators
\ba\label{14Aa}
L_1&=&diag(1,1,-1,-1)~,~L_2=diag(1,-1,1,-1),\nn\\
L_3&=&diag(1,-1,-1,1).
\ea
The expectation values $\rho_1,\rho_2,\rho_3$ determine the diagonal elements of the density matrix
\ba\label{14Ba}
\rho_{11}&=&\frac14(1+\rho_1+\rho_2+\rho_3)~,~\rho_{22}=\frac14(1+\rho_1-\rho_2-\rho_3),
\nn\\
\rho_{33}&=&\frac14(1-\rho_1+\rho_2-\rho_3)~,~\rho_{44}=\frac14(1-\rho_1-\rho_2+\rho_3).\nn\\
\ea
While the purity constraint $P\leq 3$ assures $\rho_{\alpha\alpha}\leq 1$, the positivity condition $\rho_{\alpha\alpha}\geq 0$, which is necessary for a valid density matrix, is not obeyed automatically. (For example, $\rho_1=\rho_2=\rho_3=-1~,~\rho_{k\geq 4}=0$, obeys $P=3$ but would imply $\rho_{11}=-1/2.)$ We will therefore extend the purity constraint and impose further conditions on the allowed values of $\rho_k$ (beyond the condition $\sum_k\rho_k\rho_k\leq 3$) that will guarantee $\rho_{\alpha\alpha}\geq 0$. The explicit form of this positivity constraint will not be important for the present paper and we refer for a detailed discussion to \cite{CWAA}, \cite{CW1}. In short, we require that the eigenvalues of the density matrix \eqref{10a} should all be positive or zero. For a pure state this implies $\rho_k=f_k$, with $f_k$ parameterizing the homogeneous space $SU(M)/SU(M-1)\times U(1)$, normalized with $\sum_kf_kf_k=M-1$.

\newpage\noindent
{\bf 4. Classical particle}

The specific properties of the ``two-bit-particle'' depend on the interpretation of the basis observables $A^{(k)}$. For $M=4$, we could represent a {\em classical} particle by associating the location $L$ and the motion $M$ to two commuting operators, for example
\be\label{12}
L=L_1=\textup{diag}(1,1,-1,-1)~,~M=L_2=\textup{diag}(1,-1,1-1).
\ee
Thus $L$ and $M$ can have simultaneously sharp values. If no other observables are considered, only the values of $\rho_1$ and $\rho_2$ are needed for the computation of $\kl L\kr$ and $\kl M\kr$ in the ensemble. A pure classical state has $|\rho_1|=|\rho_2|=1$. We may further consider an observable associated to the third diagonal generator $L_3=L_1L_2=$diag$(1,-1,-1,1)$. It measures the correlation between motion and location, $LM$. This observable takes the value $+1$ for a particle located at $r$ moving clockwise or a particle located at $(l)$ moving anticlockwise, while $LM=-1$ holds if $(r)$ and $(l)$ are exchanged. The most general statistical state of a two-bit classical particle is characterized by three numbers.
\be\label{13}
\rho_1=\kl L\kr~,~\rho_2=\kl M\kr~,~\rho_3=\kl LM\kr.
\ee

For the example $\rho_1=\rho_2=0~,~\rho_3=1~,~\kl LM \kr=1$, one does not know if the particle is located left or right or if it moves clockwise or anticlockwise. However, we know it moves ``downwards'', either clockwise on the right or anticlockwise on the left. As long as only the observables $L,M$ and $LM$ are used as particle observables and $\rho_1,\rho_2,\rho_3$ describe the state of the particle, we deal now with complete statistics, for which the measurement correlation $\kl LM\kr_m$ is expressed by the expectation value $\kl LM\kr=\rho_3$, which is directly connected to the joint probabilities. The characteristic feature of a classical particle is that only diagonal operators are used for a description of the observables, and that the observables include both location and motion observables. 

\medskip\noindent
{\bf 5. Quantum particle}

In contrast, for a description of a two-bit {\em quantum} particle the characterization of the particle state also uses information from observables corresponding to off-diagonal operators. We may now employ the three diagonal generators $L_1,L_2,L_3$ for a refined characterization of the location using four cells. (Imagine in our intuitive example that we place two further detectors at the location $x=\pi$ and $x=-\pi$.) We may label the cells with $\alpha=1\dots 4$ and ``central locations'' at $x(\alpha=1)=3\pi/4~,~x(\alpha=2)=\pi/4~,~x(\alpha=3)=-\pi/4~,~x(\alpha=4)=-3\pi/4$. The particle numbers $N^{(\alpha)}$ obey 
\ba\label{14}
\hat N^{(1)}&=&\frac14 (1+L_1+L_2+L_3),\nn\\
\hat N^{(2)}&=&\frac14(1+L_1-L_2-L_3),\nn\\
\hat N^{(3)}&=&\frac14(1-L_1+L_2-L_3),\nn\\
\hat N^{(4)}&=&\frac14(1-L_1-L_2+L_3),\nn\\
\ea
and the location of the particle is given by $\hat X$ according to eq. \eqref{6}, with eq. \eqref{11} equivalent to eq. \eqref{2}. The observable $L_1$ takes positive values if the particle is on the right $(x>0)$,  and negative values for a particle on the left $(x<0)$, according to
\be\label{14A}
L_1=\hat N^{(1)}+\hat N^{(2)}-\hat N^{(3)}-\hat N^{(4)}.
\ee
Similarly, $L_3$ is positive (negative) for a particle in the lower half, $|x|>\pi/2$ (upper half, $|x|<\pi/2)$,
\be\label{14B}
L_3=\hat N^{(1)}-\hat N^{(2)}-\hat N^{(3)}+\hat N^{(4)},
\ee
and $L_2$ is positive or negative
\be\label{14C}
L_2=\hat N^{(1)}-\hat N^{(2)}+\hat N^{(3)}-\hat N^{(4)}
\ee
in one or the other of the two complementary diagonal regions of the circle. 

The off-diagonal generators $L_4\dots L_{15}$ may be split into real generators $(L^*_k=L_k)$, namely $L_4=1\otimes\tau_1~,~L_6=\tau_3\otimes\tau_1$ , $L_8=\tau_1\otimes 1~,~L_{10}=\tau_1\otimes\tau_3~,~L_{12}=\tau_1\otimes\tau_1~,~L_{14}=-\tau_2\otimes\tau_2$ and imaginary generators $(L^*_k=-L_k)$, namely $L_5=1\otimes\tau_2$, $L_7=\tau_3\otimes\tau_2~,~L_9=\tau_2\otimes 1~,~L_{11}=\tau_2\otimes\tau_3~,~L_{13}=\tau_1\otimes\tau_2$ , $L_{15}=\tau_2\otimes\tau_1$. (In this language one has $L_1=\tau_3\otimes 1~,~L_2=1\otimes\tau_3~,~L_3=\tau_3\otimes\tau_3$.) We may employ the six imaginary generators for generalized motion observables. One motion observable $\hat M_1$ can be associated with $-L_5$, such that $L_5=1$ for anticlockwise motion, and $L_5=-1$ for clockwise motion. We observe that $L_5$ and $L_1$ commute such that there can be states where they have simultaneously sharp values. Indeed, a state with $\rho_1=\kl L_1\kr=1$, $\rho_5=\kl L_5\kr=-1$ characterizes the particle as being in the right half of the circle {\em and} moving clockwise. In this special case we can recover  a classical particle where a location and a motion observable are simultaneously measurable (even though with reduced precision, since only one bit can be used for the location measurement). Indeed, we may employ a change of basis for the generators $L_k$ by unitary transformations (with a simultaneous transformation of $\rho$ such that $\tr(\rho\hat A)$ remains invariant). This can be used in order to transform the triplet of generators $(L_1,-L_5,-L_7)$ into $(L_1,L_2,L_3)$, and we see the equivalence with the previous description of a classical particle if no information on further observables is used. This allows an interpretation of the observable associated to $L_7$. For $L_7=1$ the particle is either on the right moving anticlockwise or on the left moving clockwise (the particle moves ``upwards''), while for $L_7=-1$ it either moves clockwise on the right or anticlockwise on the left (``downwards'' motion). 

The observable $M_1$ associated to $\hat M_1=-L_5$ is not the only motion observable. A different motion observable is associated to $\hat M_2=L_{13}$, with positive $L_{13}$ describing clockwise motion. This operator commutes with $L_3$, such that we can now simultaneously answer the question if the particle is in the lower half $(L_3=1)$ or in the upper half $(L_3=-1)$, and if it is moving clockwise $(L_{13}=1)$ or anticlockwise $L_{13}=-1$. Again, the triplet of operators $(L_3,L_{13},L_{15}=L_3L_{13})$ is equivalent by unitary transformations to $(L_1,L_2,L_3)$ and we can describe a classical particle. The third set of motion type observables, $-L_9$ and $-L_{11}=-L_2L_9$, commutes with $L_2$. Its interpretation is less intuitive since it involves disconnected regions for the location in the ring. 

The two motion operators $\hat M_1$ and $\hat M_2$ commute (with $\hat M_1\hat M_2=-L_8)$. We may define an angular momentum observable
\be\label{15}
\hat M=\frac12(\hat M_1+\hat M_2)=\frac12(L_{13}-L_5)=\frac i2
\left(\begin{array}{cccc}
0&1&0&-1\\
-1&0&1&0\\
0&-1&0&1\\
1&0&-1&0
\end{array}\right)
\ee
This operator has three different eigenvalues, $m=0,\pm 1$, as characteristic for angular momentum in quantum mechanics in appropriate units. We observe that $\hat M$ does not commute with anyone of the location operators $L_1,L_2,L_3$ or with $\hat X$.

Pure states correspond to classical ensembles for which the density matrix obeys $\rho^2=\rho$. This requires a maximal purity $P=3$. Any pure state density matrix can be written as
\be\label{16}
\rho=U\hat\rho_1U^\dagger~,~\hat\rho_1=\textup{diag } (1,0,0,0),
\ee
with $U$ a suitable unitary matrix $U^\dagger U=1$. For pure states we may introduce the quantum mechanical wave function as a normalized complex $M$-component vector $\psi,\psi^\dagger\psi=1$, according to 
\be\label{17}
\psi_\alpha=U\ab\hat\psi_1~,~(\hat\psi_n)_\alpha=\delta_{n\alpha}~,
~\rho\ab=\psi_\alpha\psi^*_\beta,
\ee
such that the expectation value of an observable $A$ can be computed using the associated operator $\hat A$ with the standard quantum mechanical rule
\be\label{18}
\kl A \kr =\psi^\dagger\hat A\psi.
\ee
The eigenstates of the location operator $\hat X$ are the eigenstates of the local particle numbers $\hat\psi_n$, with eigenvalues $x_n=(5-2n)\pi/4$. For all these ``localized states'' the average of the angular momentum observable vanishes, $\psi^\dagger\hat M\psi=0$. 

The eigenstates of the angular momentum operator are given by
\be\label{19}
\psi_\alpha=\frac12\exp[i\tilde m x(\alpha)]~,~x(\alpha)=\frac{(5-2\alpha)\pi}{4},
\ee
with $\tilde m=0,~\pm 1,~\pm 2$. We notice that $\tilde m=+2$ and $\tilde m=-2$ leads to the same state 
$\psi=\pm\frac i2 (1,-1,1,-1)$ up to a minus sign. It corresponds to a zero eigenvalue of $\hat M$. For $|\tilde m|\leq 1$ one has $\tilde m=m$. For the angular momentum eigenstates one finds
\be\label{20}
\kl N^{(\alpha)}\kr=\psi^\dagger\hat N^{(\alpha)}\psi=\frac14~,~\kl X\kr=\psi^\dagger\hat X\psi=0.
\ee
It is amazing how many characteristic features of quantum mechanics are already visible for our simple ``two bit particle''!

Indeed, we can now characterize a two-bit quantum particle by a subsystem with the following properties: (i) All particle observables have a spectrum of at most four different values, corresponding to two yes/no questions. (ii) The state of the subsystem cannot be fully characterized by the expectation values for three ``commuting observables'', for which joint probabilities are available. It needs the specification of expectation values of further observables. (iii) The statistical system describing the particle alone is incomplete. Joint probabilities are not available for all pairs of observables. (iv) The location and motion observables use a maximal resolution consistent with two bits. In consequence, they cannot be associated to commuting operators. 

Finally, we may describe the classical particle within the quantum mechanical formalism as ``classical particle eigenstates'' which have simultaneously sharp values of $L_1$ and $\hat M_1$ (and therefore also of $L_7$). The eigenstates of $L_1=+1$  obey $\psi=(\psi_1,\psi_2,0,0)$. The simultaneous eigenstates with $M_1=\pm 1$ read $\psi=(1,\mp i,0,0)/\sqrt{2}$. (For the eigenstates with $L_1=-1$ one exchanges the upper two with the lower two components.) In a quantum mechanical language, the classical particle states are entangled states between the location eigenstates $\psi_1$ and $\psi_2$. 

\medskip\noindent
{\bf 6. Classical statistical ensemble for two-bit-particle}

There are many possible classical statistical ensembles that realize a two-bit-particle. They typically differ in the properties of the environment, while they lead to identical results for the subsystem which characterizes the two-bit-particle. As a first explicit example for a classical statistical ensemble we may consider $2^{15}$ classical states $\tau$ which are characetrized by ordered sequences $\{\sigma_k\}$ of fifteen discrete variables $\sigma_k=\pm 1$. We specify the state of the subsystem by the expectation values of the fifteen observables $\sigma_k$, 
\be\label{28A}
\rho_j=\kl\sigma_j\kr=\sum_{\{\sigma_k\}}\sigma_j p\big(\{\sigma_k\}\big).
\ee
For the probabilities $p_\tau\equiv p\big(\{\sigma_k\}\big)$ of the classical statistical ensemble we choose
\be\label{28B}
p\big(\{\sigma_k\}\big)=p_s\big(\{\sigma_k\}\big)+\delta p_e\big(\{\sigma_k\}\big)
\ee
with
\be\label{28C}
p_s\big(\{\sigma_k\}\big)=2^{-15}\prod^{15}_{k=1}(1+\rho_k\sigma_k).
\ee
The part $\delta p_e$ obeys
\be\label{28D}
\sum_{\{\sigma_k\}}\sigma_j\delta p_e\big(\{\sigma_k\}\big)=0~,~
\sum_{\{\sigma_k\}}\delta p_e\big(\{\sigma_k\}\big)=0,
\ee
such that it only matters for the environment, without influencing the state of the subsystem. This example can be easily generalized by extending the number of classical states, using an additional index $\zeta$ with $\tau=\big(\{\sigma_k\},\zeta\big)$. The probability $p_s\big(\{\sigma_k\}\big)$ is multiplied by $\bar p_s(\zeta)$ obeying $\sum_\zeta\bar p_s(\zeta)=1$, and $\delta p_e$ depends on $\zeta$ in addition to $\{\sigma_k\}$. The sums in eq. \eqref{28A}, \eqref{28D} became now sums over $\{\sigma_k\}$ and $\zeta$. For example, the sequences of hits discussed in sect. \ref{quantumandclassical} can be described in this way.

We may associate to each classical observable $\sigma_k$ the operator $L_k$ and define the density matrix
\be\label{28E}
\rho=\frac14 (1+\rho_k L_k).
\ee
This guarantees the quantum rule for the expectation values 
\be\label{28F}
\kl A\kr=\text{tr}(\hat A\rho)
\ee
if $A=\sigma_k$ and $\hat A=L_k$. We constrain the ``purity'' $P=\sum_k\rho_k\rho_k\leq 3$ and impose further on the allowed values of $\rho_k$ the positivity constraint that all eigenvalues of $\rho$ should be positive or vanish. Then $\rho$ in eq. \eqref{28E} has all the properties of a density matrix for a four-state quantum system. A unitary time evolution of the density matrix $\rho$ can be achieved by a suitable time evolution of the classical probability distribution \eqref{28B}, as realized by an appropriate rotation among the $\rho_k$, while the evolution of $\delta p_e$ is arbitrary as long as the constraints \eqref{28D} are respected.

One may realize the occupation numbers \eqref{14} in the classical statistical ensemble by appropriate sums of $\sigma_1,\sigma_2$ and $\sigma_3$. In general, these sums could take the values $\left(1,\frac12,0,-\frac12\right)$, instead of the allowed values $(1,0)$ for the occupation numbers. However, we may impose on the allowed classical probability distributions $p_\tau$ a further restriction such that $\sigma_1,\sigma_2$ and $\sigma_3$ form a ``comeasurable bit chain'' \cite{CWAA}. This means that their expectation values obey the relation
\ba\label{28G}
\kl\sigma_1\cdot\sigma_3\kr&=&\kl\sigma_2\kr~,~\kl\sigma_1\cdot\sigma_2\kr=\kl\sigma_3\kr~,~
\kl\sigma_2\cdot\sigma_3\kr=\kl\sigma_1\kr,\nonumber\\
\kl\sigma_1\cdot\sigma_2\cdot\sigma_3\kr&=&1.
\ea
Comeasurable bit chains are possible if the associated quantum operators commute. The relations \eqref{28G} are then realized also on the operator level
\be\label{28H}
L_1L_3=L_2~,~L_1L_2=L_3~,~L_2L_3=L_1~,~L_1L_2L_3=1.
\ee
The relations \eqref{28G} imply that the observables $A^{(j)}$, to which the operators $\hat N^{(j)}$ are associated, behave as projectors
\be\label{28I}
\kl (A^{(j)})^p\kr=\kl A^{(j)}\kr.
\ee
This may be verified explicitely for $p=2,3$, and the use of the identity
\be\label{28J}
(A^{(j)})^4=(A^{(j)})^3+\frac14(A^{(j)})^2-\frac14(A^{(j)}).
\ee
The relation \eqref{28I} implies that the probabilities for finding for the sum the values $\pm\frac12$ vanishes, such that the spectrum of the observables $A^{(j)}$ contains indeed only the values $1$ and $0$. Further comeasurable bit chains can be constructed for the motion observables $(-\sigma_5,\sigma_{13},-\sigma_8)$, in complete analogy to $(\sigma_1,\sigma_2,\sigma_3)$. We can also implement a comeasurable bit chain $(\sigma_1,-\sigma_5,-\sigma_7)$ for the classical particle. For a two-bit-particle or four-state quantum mechanics the maximal numbers of members of a bit chain is three. 

Already on the level of two-bit subsystems we have constructed classical statistical ensembles that can describe a quantum particle as well as a classical particle. The ``nature'' of the particle depends on what observables are used for the description. In our $2$-bit example, the angular momentum operator for a quantum particle is $\hat M$ (eq. \eqref{15}), while for a classical particle one uses $\hat M_1=-L_5$. For a classical particle, typically not all of the information contained in the density matrix $\rho$ is used - with the exception of states where $\rho_k=0,k\neq 1,5,7$. While the location infomation of a quantum particle uses the three expectation values $\rho_1,\rho_2,\rho_3$, only $\rho_1$ is used for a classical particle. In the next sections we will generalize these findings by considering a continuum limit where the number of bits goes to infinity, thereby increasing the resolution of the location observable $\hat X$ and the range of the angular momentum $\hat M$.

\section{Particle motion}
\label{particlemotion}
In this section we extend the discussion from the ``one-bit-particle'' and ``two-bit-particle'', where all observables have a discrete spectrum, to quantum and classical particles described by continuous position and momentum observables. For the quantum particle we will obtain Heisenberg's uncertainty relation and non-commuting position and momentum operators. In contrast, for a classical particle position and momentum can both be measured sharply and the associated operators commute.

\medskip\noindent
{\bf 1. Continuous observables}

Let us next increase the number of bits $Q$ and the corresponding number of quantum states $M=2^Q$. We finally will be interested in the continuum limit $M\to\infty$. We now consider a basis of $M^2-1$ two-level observables $\A$ for describing the state of the particle by eq. \eqref{8a}. Equations (\ref{9a}-\ref{11}) have been already formulated for general $M$ (now $P \leq M-1)$ and we keep the same normalization for the $SU(M)$ generators $L_k$. Our quantum mechanical formalism holds for general $M$. For the particle numbers $\hat N^{(\alpha)}$ we use eq. \eqref{5}, with $\alpha=1\dots M$. Defining
\be\label{21}
x(\alpha)=(M+1-2\alpha)\frac\pi M
\ee
we infer the ``quantum location operator'' $\hat X$ as
\be\label{22}
\hat X=\sum_\alpha x(\alpha)\hat N^{(\alpha)}.
\ee
Thus $x$ remains an angular variable on the circle, $-\pi<x<\pi$, but the resolution increases due to the higher number of possible eigenstates.

For $M\to\infty$ the location becomes a continuous variable. We may switch to a continuum notation where $\psi_\alpha$ is replaced  by the continuous wave function $\psi(x)$. The density matrix becomes a function of two coordinates, $\rho(x,y)$. For the special case of a pure state it reads $\rho(x,y)=\psi(x)\psi^*(y)$. Similarly, the operators become matrices in position space, $\hat A(x,y)$. In particular, the particle number at location $z$ reads
\be\label{23}
\hat N^{(z)}(x,y)=\delta(x-z)\delta(y-z).
\ee
The quantum position operator takes the form
\be\label{24}
\hat X(x,y)=\int dz~ z\hat N(z)(x,y)=x\delta(x-y),
\ee
with expectation value
\be\label{25}
\kl X\kr =\int_x\int_y\hat X(x,y)\rho(y,x)=\int_x x\rho(x,x).
\ee
We can interprete $\rho(x,x)$ as the probability $w(x)$ to find the particle at location $x$. It obeys $0\leq w(x)\leq 1$ and the condition $\tr \rho=1$ assures the proper normalization $\int_x w(x)=1$. For a pure state we have the usual rule
\be\label{26}
\kl X\kr =\int_x\psi^*(x)x\psi(x)~,~w(x)=\psi^*(x)\psi(x).
\ee

\medskip\noindent
{\bf 2. Momentum and angular momentum for 

\hspace{0.15cm}quantum particle}

Consider next the angular momentum operator $\hat M$ for a general number of quantum states $M$. (There should be no confusion between the similar symbols.) It reads 
\be\label{27}
\hat M\ab =\frac i{\cal N}
\sum^M_{\gamma=1}(\delta_{\alpha,\gamma}\delta_{\beta,\gamma+1}
-\delta_{\alpha,\gamma+1}\delta_{\beta,\gamma}).
\ee
Here $\gamma$ is considered as an index modulo $M$, i.e. $\gamma=M+1$ corresponds to $\gamma=1$. The normalization ${\cal N}$ depends on the units for $\hat M$. We choose
\be\label{28}
{\cal N}=2\sin\frac{2\pi}{M}
\ee
such that 
\be\label{29}
\psi_\alpha=\frac{1}{\sqrt{M}}\exp ix(\alpha)
\ee
is an eigenstate of $\hat M$ with eigenvalue $m=1$. One finds for the eigenvalue $\tilde m(m)$ of $\hat M$, with eigenfunctions
\be\label{30}
\psi^{(m)}_\alpha=\frac{1}{M}\exp \big[imx(\alpha)\big]~,~-\frac{M}{2}< m<\frac M2
\ee
the spectrum
\be\label{31}
\tilde m(m)=\frac{\sin(2\pi m/M)}{\sin(2\pi/M)}.
\ee

For $M\to\infty$ the small values of $\tilde m$ become the usual integer values of angular momentum, $\tilde m(m)\to m$. For $m=M/4$ the function $\tilde m(m)$ reaches its maximum $\tilde m_{\textup{max}}=1/\sin(2\pi/M)\approx M/2\pi$, and then decreases towards zero for $m\to M/2$. We see that the range of possible angular momenta increases $\sim M$, in correspondence with the increased angular resolution, $\Delta x\sim 1/M$. The appearance of small eigenvalues $\tilde m$ for $|m|\to M/2$ is an artefact of the lattice formulation (similar to the fermion doubling in lattice gauge theories). They may be removed towards large values by the use of an improved angular momentum operator. We will assume that this is done and consider for the continuum limit only smooth functions with $|m|\ll M/4$. In the continuum limit we find integer eigenvalues of angular momentum. The continuum version of $\hat M$ is a differential operator
\be\label{32}
\hat M(x,y)=-i\delta(x-y)\frac{\partial}{\partial y}.
\ee
If we further restore standard units by multiplying $\hat M$ with $\hbar$ (so far we have used $\hbar=1$) one recovers the well known commutation relation between an angular position operator $\hat X$ and the angular momentum operator $\hat M$
\be\label{33}
[\hat X,\hat M]=i\hbar\delta(x-y).
\ee

The continuum limit can also be taken in a different way. While increasing the resolution we may simultaneously increase the ``volume'' of the circle (i.e. the circumference) to infinity, and consider only a finite region in this infinite volume. This essentially amounts to a change of units and to a restriction of observables to a shrinking region in $x$. We may choose some length unit and perform a rescaling $x(\alpha)=2\pi x'/l$, such that the new location variable extends from $-l/2\leq x'\leq l/2$, with $l=Ma$ the volume and $a$ the lattice distance. With $\hat X'=(l/2\pi)\hat X~,~\hat P=(2\pi\hbar/l)\hat M$, the new variables $\hat X,\hat M'$ read
\be\label{34}
\hat X'=x'\delta(x'-y')~,~\hat P=-i\hbar\delta(x'-y')\frac{\partial}{\partial y'}.
\ee
We interprete now $\hat P$ as momentum - in one dimension there is no difference between momentum in a periodic volume and angular momentum (up to units). The relation \eqref{33} becomes the well known Heisenberg's uncertainty relation between position and momentum operators (we drop the primes on $\hat X,x$ from now on)
\be\label{35}
[\hat X,\hat P]=i\hbar\delta(x-y).
\ee

For a finite $l$ the momenta remain discrete, with $p=2\pi /l~($ from here on we use again $\hbar=1)$. A scaling $l\sim \sqrt{M}, a\sim 1/\sqrt{M}$, however, leads for $M\to \infty$ to continuous momentum and location variables. Restricting the discussion to a fixed finite range $\Delta x$, with $\Delta x/l\to 0$ for $l\to \infty$, the periodicity of $x$ can be neglected. We end with the standard setting for a particle  in quantum mechanics. In this version the generalization to three dimensions is straightforward.

\medskip\noindent
{\bf 3. Commuting position and momentum for 

\hspace{0.15cm}classical particle}

What happens for $M\to\infty$ with the ``classical particle state'' that we have found in our quantum mechanical formalism for $M=4$? It is indeed straightforward to generalize for large $M$ the notion of a classical particle where both location and momentum can be measured. For this purpose we use half of the yes/no questions for the location, and the other half for the motion. (We assume here $Q/2$ to be integer.) We can construct a coarse grained location operator $\hat X_{\sqrt{M}}$ as a $\sqrt{M}\times\sqrt{M}$ matrix $(\sqrt{M}=2^{Q/2})$. It follows the same rules as the previous quantum location operator $\hat X_M$, but with less resolution since only $Q/2$ bits can be used. Similarly, we construct a coarse grained momentum operator $\hat P_{\sqrt{M}}$. The classical location and momentum operators $\hat X_{cl}$ and $\hat P_{cl}$ are given by $M\times M$ matrices
\be\label{36}
\hat X_{cl}=\hat X_{\sqrt{M}}\otimes {\mathbbm 1}_{\sqrt{M}}~,~\hat P_{cl}={\mathbbm 1}_{\sqrt{M}}\otimes\hat P_{\sqrt{M}}.
\ee
Obviously, classical location and momentum commute
\be\label{37}
[\hat X_{cl},\hat P_{cl}]=0.
\ee

This construction generalizes the classical location and motion for $M=4$. Indeed $\hat X_2=\tau_3$ and $\hat P_2=-\tau_2$ reproduce $\hat X_{cl}=\tau_3\otimes 1=L_1~,~\hat P_{cl}=1\otimes(-\tau_2)=-L_5$. In the limit $M\to\infty$ both $\hat X_{cl}$ and $\hat P_{cl}$ become continuous operators. This demonstrates clearly that our setting of describing a particle as a subsystem of a classical statistical ensemble can account for both quantum and classical particles!

There are many different ways of taking the limit $M\to\infty$ which result in the classical commutation relation \eqref{37}, and similar for the quantum commutator \eqref{35}. At the end, only the commutation relation matters for the distinction between quantum and classical particles. As an example, we may represent the location observable for both a classical and a quantum particle in the $\sqrt{M}$-dimensional subspace
\be\label{43A}
\hat X_Q=\hat X_{cl}=\hat X_{\sqrt{M}}\otimes {\mathbbm 1}_{\sqrt{M}}.
\ee
The momentum operators can then be represented as acting in different subspaces
\be\label{43B}
\hat P_Q=\hat P_{\sqrt{M}}\otimes {\mathbbm 1}~,~\hat P_{cl}={\mathbbm 1}\otimes \hat P_{\sqrt{M}}.
\ee
In other words, the $M$-dimensional position space can be parameterized by a pair $(x_1,x_2)$, with representations
\be\label{43C}
\hat X_Q=\hat X_{cl}=x_1~,~\hat P_Q=-i\hbar\frac{\partial}{\partial x_1}~,~
\hat P_{cl}=-i\hbar\frac{\partial}{\partial x_2}.
\ee
In this representation the coordinate $x_2$ becomes irrelevant for the observable describing a quantum particle.

\section{Particle correlation for quantum particle}
\label{particlecorrelation}
So far we have only discussed the diagonal and the imaginary part of the density 
matrix for quantum particles. In our basis they are related in a general sense to the location and motion of the particle. The real off-diagonal part of the operators and the associated part of the density matrix describes a type of single particle correlation. It generalizes the observable $C$ for the one bit system which takes the value $+1$ if the particle has a tendency to be both right and left (correlation), and $-1$ if the tendency is towards mutual exclusion of the two locations. Similarly, for the two bit system a density matrix
\be\label{ZA}
\rho=\frac12\left(\begin{array}{llll}
1,&1,&0,&0\\
1,&1,&0,&0\\
0,&0,&0,&0\\
0,&0,&0,&0\\
\end{array}\right),
\ee
indicates a tendency that the particle is simultaneously in the first quarter of the circle $(x=3\pi/4)$ and in the second $(x=\pi/4)$. One finds $\kl N^{(1)}\kr=\kl N^{(2)}\kr=1/2$, while $\kl N^{(3)}\kr=\kl N^{(4)}\kr=0$. However, the state is not a mixed state with equal probability of the particle to be in one of the first two quarters. Such a mixed state would correspond to a density matrix $\rho=diag(1/2,1/2,0,0)$, with $\tr\rho^2=1/2$, whereas the correlated state \eqref{ZA} is a pure state with $\rho^2=\rho$. Indeed, the density matrix \eqref{ZA} can be associated to a wave function $\psi=(\psi_1+\psi_2)/\sqrt{2}$. 

\medskip\noindent
{\bf 1. Extended wave functions as eigenstates of 

\hspace{0.15cm}correlation operators}

This can be generalized immediately to the continuum limit with infinitely many bits. Consider a wave function which corresponds to a Gaussian wave packet
\be\label{ZB}
\psi(x)=(2\pi\Delta^2)^{-\frac14}\exp\left\{-\frac{(x-\bar x)^2}{4\Delta^2}\right\}.
\ee
The corresponding density matrix
\ba\label{ZC}
\rho(x,y)&=&(2\pi\Delta^2)^{-\frac12}\exp 
\left\{-\frac{(x+y-2\bar x)^2}{8\Delta^2}\right\}\nn\\
&\times&\exp\left\{-\frac{(x-y)^2}{8\Delta^2}\right\}
\ea
shows a nonvanishing correlation between distant $x$ and $y$. For example, the operator $\hat A=\delta(x-y-2a)$ can be used in order to test the shape of the disctribution. The density matrix \eqref{ZC} leads to a nonvanishing expectation value of the associated observable
\be\label{ZD}
\kl A\kr=\int_{xy}\rho(x,y)\delta(x-y-2a)=\exp\left(-\frac{a^2}{2\Delta^2}\right),
\ee
which vanishes for large $a$ with a characteristic width given by $\Delta$

The expectation values of the location observable $X$ and its dispersion read
\be\label{ZE}
\kl X\kr=\bar x~,~\kl(X-\bar x)\kr=\Delta^2.
\ee
Thus eq. \eqref{ZC} describes a situation where physics in neighboring regions, $|x-y|/2\lesssim \Delta$, is strongly correlated. We may associate $Re\big(\rho(x,y)\big)$ with a correlation function.

The Gaussian wave function \eqref{ZB} is an eigenstate of the family of hermitean correlation operators
\ba\label{48A}
&&\hat C_\Delta(\bar x;x,y)=\frac{1}{4\pi\Delta^2}\\
&&\times \exp 
\left\{-\frac{1}{4\Delta^2}
\left[a(x-y)^2+\frac 1a
\left(\frac{x+y}{2}-\bar x\right)^2\right]\right\},\nn
\ea
obeying
\be\label{48B}
\int dy~\hat C_\Delta(\bar x;x,y)\psi(y)=c_\Delta\psi(x),
\ee
with eigenvalues
\be\label{48C}
c_\Delta=\frac{1}{1+2a}\left(\frac{a}{\pi}\right)^{1/2}\frac1\Delta.
\ee
We have normalized $\hat C_\Delta$ such that in the limit $\Delta\to 0$ it becomes the occupation number at location $\bar x$
\be\label{48D}
\lim_{\Delta\to 0}\hat C_\Delta=\hat N^{(\bar x)}=\delta(x-\bar x)\delta(y-\bar x).
\ee
However, for any $\Delta>0$ the operators $\hat C_\Delta$ are real symmetric operators of the particle correlation type. The family of operators depends on a parameter $a$. For $a\to\infty$ the correlation operator $\hat C_\Delta$ becomes proportional to the unit operator and independent of $\bar x$, while for $a\to 0$ it becomes proportional to $\delta$-distribution for the center of mass coordinate $z=(x+y)/2$, i.e. $\sim\delta(z-\bar x)$, while the relative coordinate plays no role. 

\medskip\noindent
{\bf 2. Wave packets}

As well known from quantum mechanics, a moving quantum particle can be described by a wave packet. We recapitulate here the basic properties in order to facilitate the comparison with wave packets for classical particles that will be discussed in the next section. Let us consider a wave function
\be\label{ZF}
\psi(x,t)=\int\frac{dp}{2\pi}e^{i(px-\omega(p)t)}A(p),
\ee
where we choose for definiteness a nonrelativistic free particle with $\omega(p)=p^2/2m$, and a Gaussian wave packet,
\be\label{ZG}
A(p)=\left(\frac{\Delta^2_p}{2\pi}\right)^{-\frac14}\exp \left\{-\frac{(p-\bar p)^2}{4\Delta^2_p}\right\}.
\ee
In Fourier space this yields
\be\label{50A}
\psi(p,t)=A(p)\exp \left(-i\frac{p^2t}{2m}\right).
\ee
Again, the wave packet \eqref{ZF} describes a state with non-vanishing particle correlation for $y\neq x$, similar to the static Gaussian distribution \eqref{ZB}.

The corresponding pure state density matrix reads
\ba\label{ZH}
\rho(x,y;t)&=&\sqrt{\frac{2}{\pi}}\bar\Delta_p
\exp\left\{-2\bar\Delta^2_p
\left(\frac{x+y}{2}-\frac{\bar p t}{m}\right)^2\right\}\nn\\
\times&\exp& \left\{-\frac{\bar\Delta^2_p}{2}(x-y)^2\right\}\\
\times&\exp&\left\{i(x-y)\left[\bar p+\frac{4\Delta^2_p\bar\Delta^2_p t}{m}
\left(\frac{x+y}{2}-\frac{\bar pt}{m}\right)\right]\right\}\nn
\ea
with
\be\label{ZI}
\bar\Delta_p=\Delta_p\left(1+\frac{4\Delta^4_pt^2}{m^2}\right)^{-\frac12}
\ee
In the limit $\Delta_p\to 0$ one recovers the plane wave density matrix (eigenstate of momentum)
\be\label{ZJ}
\rho_{\bar p}(x,y)\to \sqrt{\frac2\pi}\Delta_p\exp[i\bar p(x-y)],
\ee
while for $\bar\Delta_p\to\infty$ one finds the limit of a sharply located particle
\be\label{ZK}
\rho(x,y)\to \frac{\sqrt{2\pi}}{\bar\Delta_p}\delta
\left(\frac{x+y}{2}-\frac{\bar p t}{m}\right)\delta(x-y).
\ee

Using the momentum operator \eqref{34}, $\hat P(y,x)=-i\delta(y-x)\frac{\partial}{\partial x}$, the density matrix \eqref{ZH} describes a time independent momentum distribution
\ba\label{ZL}
\kl F(p)\kr&=&\int_{x,y}\big(F(\hat P)\big)(y,x)\rho(x,y;t)\\
&=&\lim_{y\to x}\int_x F(-i\partial_x)\rho(x,y;t)\nn\\
&=&\int \frac{dp}{2\pi}F(p)\rho(p,p;t)
=\int \frac{dp}{2\pi}F(p)A^2(p).\nn
\ea
Here we employ the Fourier transform of the density matrix
\ba\label{ZM}
\rho(p,q;t)&=&\int_x\int_y\rho(x,y;t)e^{-ipx}e^{iqy}\nn\\
&=&A(p)A(q)\exp\left\{\frac{it}{2m}(q^2-p^2)\right\},
\ea
and $\kl P\kr =\bar p~,~\kl (P-\bar p)^2\kr=\Delta^2_p$. 

On the other hand, functions of the location observable obey
\ba\label{ZN}
\kl F(X)\kr&=&\int_x F(x)\rho(x,x;t)\\
&=&\sqrt{\frac{2}{\pi}}\bar\Delta_p\int_x F(x)\exp
\left\{-2\bar\Delta^2_p\left(x-\frac{\bar p t}{m}\right)^2\right\}.\nn
\ea
This implies for the location and its dispersion
\be\label{ZO}
\kl X\kr=\bar x(t)=\frac{\bar p t}{m}~,~\kl\big(X-\bar x(t)\big)^2\kr=\frac{1}{4\bar\Delta^2_p}.
\ee
For $t\to\infty$ the vanishing of $\bar\Delta_p(t\to\infty)\to 0$ implies a diverging dispersion, with $\bar\Delta_p\to m/(2\Delta_pt)$ and
\ba\label{ZP}
&&\rho(x,y;t\to\infty)\to\nn\\
&&\frac{m}{\sqrt{2\pi}\Delta_pt}\exp\left\{-\frac{1}{2\Delta^2_p}
\left(\bar p-\frac{m(x+y)}{2t}\right)^2\right\}\\
&&\times\exp\left\{-\frac{1}{2\Delta^2_p}\left(\frac{m(x-y)}{2t}\right)^2\right\}\nn\\
&&\times\quad \exp\left\{i\frac{m(x+y)}{2t}(x-y)\right\}.\nn
\ea

We finally display the Wigner representation \cite{Wig} of the density matrix \eqref{ZH} (Wigner quasi-probability distribution) by using the center of mass coordinate $z=(x+y)/2$ and performing a Fourier transformation with respect to the relative coordinate $x-y$,
\ba\label{ZQ}
\rho_w(z,q;t)&=&\int d(x-y)e^{-iq(x-y)}
\rho(x,y;t)\nn\\
&=&2\exp\Bigg\{-\frac{1}{2\Delta^2_p}(q-\bar p)^2\}\Bigg\}\nn\\
&&\times \exp \Bigg\{-2\Delta^2_p\left(z-\frac{qt}{m}\right)^2\Bigg\}.
\ea
For this particular case of a free particle $\rho_w(z,q)$ constitutes a probability density in the classical phase space of location and momentum. It is real, positive, and normalized,
\be\label{ZR}
\int dz\int\frac{dq}{2\pi}\rho_w(z,q;t)=1.
\ee
Marginalizing over the center of mass distribution yields the normalized probability distribution of momenta
\be\label{ZS}
\bar\rho(q)=\int dz\rho_w(z,q)=\frac{\sqrt{2\pi}}{\Delta_p}\exp \left\{
-\frac{1}{2\Delta^2_p}(q-\bar p)^2\right\}=A^2(q),
\ee
while integrating over momenta we find the distribution of the center of mass coordinate around the average classical trajectory $\bar z=\bar pt$, 
\ba\label{ZT}
\bar\rho(z)&=&\int\frac{dq}{2\pi}\rho_w(z,q)\nn\\
&=&\sqrt{\frac{\pi}{2}}\bar\Delta_p\exp\left\{-2\bar\Delta^2_p
\left(z-\frac{\bar p t}{m}\right)^2\right\}.
\ea

We conclude that a quantum particle is typically described by nontrivial particle correlations. In the Wigner representation the density matrix for our particular wave packet for the free particle defines a probability distribution in classical phase space of location and momentum. It is well known that this property does not hold for a general density matrix of a quantum particle - the Wigner function $\bar\rho_w$ can be negative in certain regions of phase space. Furthermore, the positivity of $\bar\rho_w$ may not be preserved by the time evolution for an interacting particle. 

\section{Quantum formalism for classical particle}
\label{classicalparticle}
Let us now turn again to the possibility of realizing a probability distribution for a ``particle'' with arbitrarily accurate location and momentum. Since we can associate a quantum mechanical density matrix to this probability distribution, this constitutes a way to describe a ``classical particle'', with sharp location and momentum, within the formalism of quantum mechanics. Of course, the relevant location and momentum operators must now commute, cf. eq. \eqref{37}. We emphasize that this classical particle is not the usual classical limit of quantum mechanics. In this section we rather develop the formalism of quantum mechanics for a microscopic classical particle. 

This part does actually not need our previous discussion how quantum mechanics arises from classical statistics. It constitutes a self-consistent probabilistic description of a classical particle. The basic concepts are the probability distribution in phase space and its dynamics. The latter is formulated as a time evolution equation for the probability distribution. This replaces the notion of trajectories and Newton's laws as basic concepts for classical particles. We first discuss the Hamiltonian which leads to a time evolution which corresponds to classical trajectories for particles. In the presence of a potential, the time evolution according to the Schr\"odinger equation associated to this Hamiltonian changes the shape of a wave packet. We establish that the probability density in phase space, which obtains from a given initial wave function, describes precisely the classical probability distribution evolving according to the Liouville equation. In the traditional approach the latter follows if non-interacting particles move on classical trajectories in a potential, with a distribution of initial conditions given by the squared initial wave function. 

A new basic concept in our treatment will be the wave function for a classical particle, which shares many formal features with the wave function for a quantum particle. Important distinctions remain, however. The classical wave function depends on both position and momentum and is a real function. 

\medskip\noindent
{\bf 1. Wave function and quantum Hamiltonian for 

\hspace{0.15cm}classical particles}

The direct product structure \eqref{36} for classical location and momentum operators implies that for each location $x$ we can assign further quantum numbers. Due to the vanishing of the commutator \eqref{37} we can choose the wave function to be simultaneously an eigenfunction of $\hat X_{cl}$ and $\hat P_{cl}$. A general wave function depends then on both the variables $x$ and $p$, with
\be\label{zA}
\hat X_{cl}\psi(x,p,t)=x\psi(x,p,t)~,~\hat P_{cl}\psi(x,p,t)=p\psi(x,p,t).
\ee
The structure of quantum mechanics perfectly allows for commuting position and momentum operators, which are represented here as
\ba\label{zB}
&&\hat X_{cl}=x\delta(x-y)~,~\hat P_{cl}=p\delta(p-q),\nn\\
&&[\hat X_{cl},\hat P_{cl}]=0.
\ea
(In our normalization $\delta(p-q)$ stands in $d$ dimensions for $(2\pi)^d\delta^d(p-q)$.) For the commuting operators $\hat X_{cl}$ and $\hat P_{cl}$ we can compute expectation values of composite observables
\be\label{85A}
\kl F(\hat X_{cl},\hat P_{cl})\kr=
\int_{x,p}|\psi(x,p)|^2F(x,p).
\ee
We can therefore identify the classical probability density in phase space, $w(x,p)$, as 
\be\label{85B}
w(x,p)=|\psi(x,p)|^2.
\ee
We will see that $\psi$ can be taken real, $w=\psi^2$, such that $\psi=\pm\sqrt{w}$ is given by $w$ up to a sign. 

The wave function $\psi(x,p)$ is the central object for casting a probabilistic theory of classical particles into the quantum formalism. It obeys the generalized Schr\"odinger equation
\be\label{89A}
i\partial_t\psi(x,p)=H\psi(x,p),
\ee
with $H$ the ``quantum Hamiltonian''. For an appropriate choice of $H$ the time evolution of the probability distribution $w=|\psi|^2$ is determined by the Liouville equation, as standard for classical particles. All the usual quantum rules for the computation of expectation values of observables apply.

The difference between a classical and a quantum particle does not arise from the different formal structure between quantum mechanics and classical statistics. We have seen before how to implement a quantum particle within the conceptual framework of classical statistics, and we establish now how to describe a classical particle within the formalism of quantum mechanics. The difference between quantum or classical behavior of a particle is rather due to the different dynamics or, in other words, to different Hamiltonians which use different types of operators.

As a special limiting case we want to describe within the formalism of quantum mechanics a particle with sharp location and momentum which follows a classical trajectory. For this purpose we cannot use a Hamiltonian $H_{cl}=V(\hat X_{cl})+\hat P^2_{cl}/2m$, since in this case both $\hat X_{cl}$ and $\hat P_{cl}$ would commute with $H_{cl}$, resulting in conserved $x$ and $p$. We may, however, use a Hamiltonian which involves both $\hat P_{cl}$ and a derivative operator which corresponds to the usual momentum operator in quantum mechanics. In this section we concentrate on a Hamiltonian which reads in the $(x,p)$ representation \eqref{zA}, \eqref{zB} 
\be\label{zC}
H_L=-\frac{i\hbar}{m}p
\frac{\partial}{\partial x}+i\hbar\frac{\partial V}{\partial x}
\frac{\partial}{\partial p},
\ee
with $V=V(\hat X_{cl})$ represented as $V(x)$ and omitting the $\delta$-functions. Other choices of the Hamiltonian are possible, as the one leading to the time evolution for a quantum particle which will be discussed in sect. \ref{Quantumparticlesfrom}, or the extended Hamiltonian investigated in appendix A.

The choice of the Hamiltonian $H_L$ leads to the commutation relations
\be\label{zD}
[H_L,\hat X_{cl}]=-\frac{i\hbar}{m}\hat P_{cl}~,~[H_L,\hat P_{cl}]=i\hbar
\frac{\partial V}{\partial\hat X_{cl}}.
\ee
(We recall that $H_L$ denotes here the Hamiltonian operator in the quantum formalism and should not be confounded with the classical Hamiltonian in classical mechanics.) A standard quantum mechanical calculation yields then the time evolution for the expectation values 
$\bar x_{cl}=\kl \hat X_{cl}\kr$ and $\bar p_{cl}=\kl\hat P_{cl}\kr$, according to 
\be\label{zE}
\partial_t\bar x_{cl}=\frac{\bar p_{cl}}{m}~,~\partial_t\bar p_{cl}=-\kl\frac{\partial V}{\partial x}\kr.
\ee
These are the same equations for the expectation values as for a quantum particle. They reduce to classical trajectories if $\kl\partial V/\partial x\kr$ can be replaced by $(\partial V/\partial x)(\bar x_{cl})$. This can be realized by a proability distribution which is sharp in position space. We emphasize that the second eq. \eqref{zE} is a perfectly classical statistical equation if we describe the time evolution for a distribution of initial conditions.

\medskip\noindent
{\bf 2. Time evolution of free classical wave packet}

We will see below that the Hamiltonian $H_L$ leads to the Liouville equation for the time evolution of the probability density $w(x,p)$. It is useful, however, to understand the time evolution of the wave function $\psi(x,p)$, as arising from the standard quantum formalism. Let us first consider the case of a free particle, $\partial V/\partial x=0$. We start with an initial wave function 
\be\label{zF}
\psi(x,p)=(\Delta_x\Delta_p)^{-\frac12}\exp 
\left\{-\frac{(x-\bar x)^2}{4\Delta_x^2}\right\}\exp 
\left\{-\frac{(p-\bar p)^2}{4\Delta_p^2}\right\}
\ee
which is normalized according to 
\be\label{zG}
\int dx\frac{dp}{2\pi}\psi^*(x,p)\psi(x,p)=1.
\ee
In the limit $\Delta_x\to 0,\Delta_p\to 0$ this is a simultaneous eigenstate of location and momentum, with eigenvalues $\bar x$ and $\bar p$. The time derivative according to the Hamiltonian \eqref{zC} reads
\be\label{zH}
\partial_t\psi=-iH_L\psi=-\frac{p}{m}\frac{\partial\psi}{\partial x},
\ee
where we use now again $\hbar=1$. Inserting a wave function of the type \eqref{zF} yields
\be\label{zI}
\partial_t\psi=\frac{p(x-\bar x)}{2m\Delta_x^2}\psi.
\ee
On the other hand, if we assume constant $\Delta_x,\Delta_p$, with only $\bar x$ depending on time, we find
\be\label{zJ}
\partial_t\psi=\frac{1}{2\Delta_x^2}(x-\bar x)(\partial_t\bar x)\psi.
\ee
The expressions \eqref{zI} and \eqref{zJ} coincide if we choose $\bar x$ depending on $p$ and $t$, $\bar x(p,t)=x_0+pt/m$. For $\Delta_x\to 0~,~\Delta_p\to 0$ we can indeed describe a free particle with sharp location and momentum, moving on a classical trajectory. As mentioned before, we can interprete $\psi^*(x,p)\psi(x,p)$ as a probability for the particle in phase space $(x,p)$. Comparing the wave function \eqref{zF} with the density matrix for a quantum particle in the Wigner representation \eqref{ZQ}, we find that $\psi^*\psi$ coincides with eq. \eqref{ZQ} for $\Delta_x=1/(2\Delta_p)$. In contrast to the quantum particle, however, we can now choose both $\Delta_x$ and $\Delta_p$ to take arbitrary values - they are not restricted by Heisenberg's uncertainty relation.

One may try to generalize the free particle construction in the presence of a potential, by using in eq. \eqref{zF} for $\bar p$ a location and time dependent mean value obeying $\partial_t\bar p(x,t)=-\partial V/\partial x$. This will, however, not yield a solution of the Schr\"odinger equation
\be\label{zK}
\partial_t\psi=-\frac pm\partial_x\psi+(\partial_x V)\partial_p\psi,
\ee
due to the additional contributions involving
\be\label{zL}
\partial_x\bar p=-\frac{\partial^2 V}{\partial x^2}~,~\partial_p\bar x=\frac tm.
\ee
For constant $\Delta_x,\Delta_p$ in eq. \eqref{zF} the l.h.s. of the Schr\"odinger equation yields
\be\label{zM}
\partial_t\psi=\frac{x-\bar x}{2\Delta_x^2}
\partial_t\bar x+\frac{p-\bar p}{2\Delta_p^2}\partial_t\bar p=
\frac{p(x-\bar x)}{2m\Delta_x^2}-\frac{p-\bar p}{2\Delta_p^2}
\frac{\partial V}{\partial x},
\ee
while one finds for the r.h.s.
\ba\label{zN}
-\frac pm\partial_x\psi+
\frac{\partial V}{\partial x}\partial_p\psi=
\left(\frac{p(x-\bar x)}{2m\Delta_x^2}-
\frac{p-\bar p}{2\Delta_p^2}\frac{\partial V}{\partial x}\right)\psi+R\psi,\nn\\
\ea
with
\ba\label{zO}
R&=&-\frac{p(p-\bar p)}{2m\Delta_p^2}\partial_x\bar p
+\frac{x-\bar x}{2\Delta_x^2}\partial_x V\partial_p\bar x\nn\\
&=&\frac{p(p-\bar p)}{2m\Delta_p^2}\frac{\partial^2 V}{\partial x^2}
+\frac{x-\bar x}{2m\Delta_x^2}\frac{\partial V}{\partial x}t.
\ea
We conclude that a wave packet given by eq. \eqref{zF}, where the time evolution arises only from $\bar x(x,p,t)$ and $\bar p(x,p,t)$, obeys the Schr\"odinger equation only for $R=0$. In the presence of a potential this will, in general, not be the case. The time evolution of the classical wave function according to the Schr\"odinger equation changes the shape of $\psi$ beyond the form \eqref{zF}.

\medskip\noindent
{\bf 3.~Classical phase space distribution from 

\hspace{0.15cm}Schr\"odinger equation}

For a classical probability distribution for pointlike particles following classical trajectories in a potential, an initial Gaussian distribution will not remain Gaussian. We may start at time $t_0=0$ with a real $\psi(x,p,0)$ given by an initial Gaussian distribution. Solving the Schr\"odinger equation, the classical probability distribution $w(x,p)$ at some later time obtains from the ``wave function'' $\psi$ as given by $w=\psi^2$,
\ba\label{zQ}
\psi(x,p,t)&=&(\Delta_x\Delta_p)^{-1}
\exp \left\{-\frac{\big(x_0(x,p,t)-\bar x\big)^2}{4\Delta_x^2}\right\}\nn\\
&&\times\exp\left\{-\frac{\big(p_0(x,p,t)-\bar p\big)^2}{4\Delta_p^2}\right\}.
\ea
Here we invert the solution for the classical trajectories for given initial values $x_0,p_0$ at $t=0$, namely $x(x_0,p_0,t),p(x_0,p_0,t)$, in order to ``extrapolate back'' the initial values $x_0,p_0$ which correspond to given values of $x(t),p(t)$. This defines $x_0(x,p,t)$ and $p_0(x,p,t)$. We assume for simplicity that two different initial conditions result in two different points in phase space $\big(x(t),p(t)\big)$ for all $t$, such that the phase space trajectory is invertible everywhere. For a real initial $\psi(x_0,p_0)$ we observe that $\psi(x,p,t)$ remains real and does not change sign during the time evolution \eqref{zQ}. This may be different if we choose a Hamiltonian different from $H_L$.

We have to show that the wave function \eqref{zQ} is a solution of the Schr\"odinger equation with Hamiltonian\eqref{zC}. For the wave function \eqref{zQ} one obtains
\ba\label{zR}
\partial_t\psi=&-&
\left[\frac{x_0(x,p,t)-\bar x}{2\Delta_x^2}\partial_tx_{0_{|x,p}}\right.\nn\\
&+&\left.\frac{p_0(x,p,t)-\bar p}{2\Delta_p^2}\partial_tp_{0_{|x,p}}\right]\psi.
\ea
On the other hand, evaluating the Hamiltonian \eqref{zC} for the wave function \eqref{zQ} yields
\ba\label{zS}
&&-iH_L\psi=-\frac{p}{m}\partial_x\psi+\partial_x V\partial_p\psi=\nn\\
&&\Bigg\{\frac{p\big(x_0(x,p,t)
-\bar x\big)}{2m\Delta_x^2}\partial_xx_{0_{|p,t}}
+\frac{p\big(p_0(x,p,t)-\bar p\big)}{2m\Delta_p^2}
\partial_xp_{0_{|p,t}}\nn\\
&&\hspace{1.5cm}-\frac{\partial_xV\big(x_0(x,p,t)-\bar x\big)}{2\Delta_x^2}\partial_px_{0{|x,t}}\nn\\
&&\hspace{1.5cm}-\frac{\partial_xV\big(p_0(x,p,t)-\bar p \big)}{2\Delta_p^2}
\partial_pp_{0_{|x,t}}\Bigg\}\psi.
\ea

We next use the observation that in the absence of an explicit time dependence of the classical Hamiltonian a given initial value $(x_0,p_0)$ can be connected to the trajectory at two different times $t_1$ and $t_2$
\be\label{zT}
(x_0,p_0)(t_1,x_1,p_1)=(x_0,p_0)(t_2,x_2,p_2),
\ee
where
\be\label{zU}
x_1=x(t_1)~,~x_2=x(t_2)~,~p_1=p(t_1)~,~p_2=p(t_2)
\ee
obey for infinitesimal $t_2-t_1$ the classical evolution equation for the trajectory (at fixed $(x_0,p_0)$)
\ba\label{zV}
x_2-x_1&=&\frac{\partial x}{\partial t}(t_2-t_1)=\frac pm(t_2-t_1),\nn\\
p_2-p_1&=&\frac{\partial p}{\partial t}(t_2-t_1)=
-\frac{\partial V}{\partial x}(t_2-t_1).
\ea
Expanding the r.h.s of eq. \eqref{zT} around $x_1,p_1$ one finds
\ba\label{zW}
\partial_tx_{0_{|x,p}}&=&-\partial _xx_{0_{|p,t}}\frac{\partial x}{\partial t}
-\partial_px_{0_{|x,t}}\frac{\partial p}{\partial t}\nn\\
&=&-\frac pm\partial_xx_{0_{|p,t}}+\partial_xV\partial_px_{0_{|x,t}},\nn\\
\partial_tp_{0_{|x,p}}&=&-\partial_xp_{0_{|p,t}}
\frac{\partial x}{\partial t}-\partial_pp_{0_{|x,t}}
\frac{\partial p}{\partial t}\nn\\
&=&-\frac{p}{m}\partial_xp_{0_{|pt}}+\partial_xV\partial_pp_{0_{|x,t}}.
\ea
Inserting these relations into eq. \eqref{zR} we find eq. \eqref{zS}. Thus the wave function \eqref{zQ}, with $x_0(x,p,t)$ and $p_0(x,p,t)$ determined by the classical trajectories, obeys the Schr\"odinger equation for the Hamiltonian \eqref{zC}. 

This finding generalizes for an arbitrary wave function of the form
\be\label{zX}
\psi(x,p,t)=\psi\big(x_0(x,p,t),p_0(x,p,t)\big).
\ee
Indeed, the relation
\be\label{zY}
\partial_t\psi_{|x,p}=\frac{\partial\psi}{\partial x_0}\partial_tx_{0_{|x,p}}
+\frac{\partial\psi}{\partial p_0}\partial_tp_{0{|x,p}}
\ee
shows that $\psi$ obeys the Schr\"odinger equation if $x_0(x,p,t)~,~p_0(x,p,t)$ are determined by the classical trajectories according to eq. \eqref{zW}, using
\ba\label{Zz}
\partial_x\psi_{|p,t}&=&\frac{\partial\psi}{\partial x_0}
\partial_xx_{0_{|p,t}}+\frac{\partial\psi}{\partial p_0}\partial_xp_{0_{|p,t}},\nn\\
\partial_p\psi_{|x,t}&=&\frac{\partial\psi}{\partial x_0}\partial_px_{0_{|x,t}}
+\frac{\partial\psi}{\partial p_0}\partial_pp_{0_{|x,t}}.
\ea
In consequence, the quantum mechanical probability $(\psi^*\psi)(t,x,p)$ for finding a particle at time $t$ at the location $x$ with momentum $p$ is exactly the same as the classical probability for pointlike particles that follow a classical trajectory each. This holds provided that at some time $t_0$ the initial classical probability distribution for the positions of the classical particles in phase space is given by $(\psi^*\psi)(t_0,x_0,p_0)$. 

In summary, we have established a quantum mechanical description for classical particles. Using a Hilbert space with commuting position and momentum operators $\hat X_{cl},\hat P_{cl}$, and the Hamiltonian \eqref{zC}, we find that states with simultaneously sharp location and momentum are possible. Furthermore, the probability distribution in phase space evolves exactly as for classical particles following classical trajectories. As long as we concentrate on ``commuting observables'', which can be written as functions of $x$ and $p$, as the energy $E=V(x)+p^2/2m$, there is a one to one correspondence between our quantum mechanical model and the usual classical description of the evolution of distributions in phase space. Indeed, since ${\hat L}=iH_L$ is a real operator containing only first derivatives, we can directly translate the Schr\"odinger equation for $\psi$ to a similar time evolution equation for the probability density in phase space
\be\label{Z1}
\partial_tw(x,p)=-iH_Lw(x,p)=-{\hat L}w(x,p).
\ee
This is the Liouville equation.

\medskip\noindent
{\bf 4. Quantum observables for classical particles}

The general solution of the Schr\"odinger equation for $\psi$ is given by
\be\label{ZZ}
\psi(x,p,t)=\psi_0\big(x_0(x,p,t),p_0(x,p,t)\big),
\ee
with $\psi_0(x_0,p_0)$ the initial wave function at some time $t_0$, and $\big(x_0(x,p,t),p_0(x,p,t)$ determined by ``following back'' the classical trajectories, as discussed above. In general, $\psi$ could be a complex normalized vector in Hilbert space. Since the evolution equation $\partial_t\psi=-{\hat L}\psi$ involves only real operators, the real and imaginary part of $\psi$ do not get mixed during the evolution. For a real initial wave function $\psi_0(x,p)$ the wave function $\psi(x,p,t)$ remains real for all $t$. In general, we may write $\psi$ in terms of the probability density $w(x,p,t)$ and a phase $\alpha(x,p,t)$, 
\be\label{Z2}
\psi(x,p,t)=w^{1/2}(x,p,t)\exp\{i\alpha(x,p,t)\}.
\ee
Only $w^{1/2}(x,p,t)$ is needed for the computation of expectation values and correlations of arbitrary observables of the type $A(x,p)$. Thus for all questions in classical physics the phase $\alpha$ is redundant. It only appears in the off-diagonal elements of the density matrix $\rho(x,y,p,q)=\psi(x,p)\psi^*(y,q)$, while it drops out from the diagonal elements $\rho(x,x,p,p)=w(x,p)$. Only the diagonal elements of $\rho$ influence the expectation values of diagonal operators $\hat A=A(x,p)\delta(x-y)\delta(p-q)$. For pure states we find for the modulus of the elements of the density matrix
\be\label{Z3}
|\rho(x,p,y,q)|^2=w(x,p)w(y,q).
\ee
Again the phase $\alpha$ plays no role.

This changes if we ask further ``quantum mechanical questions'' that do not appear in classical physics. For example, $H_L$ is a hermitean operator and we may want to compute its expectation value. For the general ansatz \eqref{Z2} it depends on the phase $\alpha(x,p)$
\be\label{Z4}
\kl H_L\kr=\int_{x,p}w\left(\frac{p}{m}\partial_x\alpha-\frac{\partial V}{\partial x}
\partial_p\alpha\right)
\ee
and differs from the average energy $\kl E\kr=\kl p^2/2m+V(x)\kr$. The additional information contained in the phase of the wave function would then be needed for the computation of expectation values of off-diagonal operators. We will, however, mainly restrict the discussion to a real wave function $\psi(x,p)$. Then the phase can only take two values $0$ and $\pi$, according to the sign of $\psi$. Instead of using $w$ and $\alpha$ it is more convenient to use directly the real function $\psi$. For real $\psi$ one finds $\kl H_L\kr=0$.

We can generalize our discussion of a ``real time evolution'' for an arbitrary antisymmetric and purely imaginary Hamiltonian. One can always define ``classical subsystems'' by restricting the observables to a set which corresponds to mutually commuting operators. In a basis where these operators are diagonal the phase of the wave function becomes irrelevant. (Note that the phase depends on quantum numbers as $(x,p)$, such that we deal with different phases for each $(x,p)$. This should not be confounded with an overall global phase of the wave function, which is always irrelevant in quantum mechanics.) Under this angle the typical quantum mechanical features are connected to the off-diagonal operators. Their expectation values may depend on the additional phase information. Still, a real time evolution always allows a setting with real $\psi$, such that the phase information concerns only the sign of $\psi$. The situation is different for Hamiltonians which contain a part which is a real function or differential operator. We discuss an example of such an ``extended Hamiltonian'' in app. A. In the following we restrict the discussion to a real wave function $\psi(x,p)$ and purely imaginary $H$. 

Let us next consider the hermitean operators
\be\label{113A}
\hat P_s=-i\partial_x~,~\hat X_s=i\partial_p
\ee
which obey
\ba\label{117A}
[\hat P_s,\hat X_{cl}]&=&-i~,~[\hat X_{s}~,~\hat P_{cl}]=i~,~[\hat X_{s},\hat P_s]=0,\nn\\
~[\hat P_s,\hat P_{cl}]&=&0~,~[\hat X_s,\hat X_{cl}]=0.\nn\\
\ea
We can write the Hamiltonian in terms of these operators
\be\label{113B}
H_L=\frac1m\hat P_{cl}\hat P_s+V'(\hat X_{cl})\hat X_s.
\ee
Both $\hat X_s$ and $\hat P_s$ are off-diagonal operators (in a basis where $\hat X_{cl}$ and $\hat P_{cl}$ are diagonal) and belong to the quantum observables. For a general complex wave function given by eq. \eqref{Z2} their expectation values depend on the phase $\alpha$. We will see how for a real wave function $\psi(x,p)$ the expectation values can actually be expressed only in terms of the probability density $w(x,p)$. 

The expectation values of $\hat X_s$ and $\hat P_s$ vanish by the absence of boundary terms
\ba\label{113C}
\kl \hat P_s\kr&=&-i\int_{x,p}\psi(x,p)\partial_x\psi(x,p)=-\frac i2\int_x\partial_xw(x,p)=0,\nn\\
\kl \hat X_s\kr&=&0.
\ea
Nevertheless, the squared operators have positive, in general nonzero, expectation values
\ba\label{113D}
\kl \hat P^2_s\kr&=&-\int_{x,p}\psi(x,p)\partial^2_x\psi(x,p)=\int_{x,p}\big (\partial_x\psi(x,p)\big)^2\nn\\
&=&\frac14\int_{x,p}w^{-1}(\partial_xw)^2=\frac14\int_{x,p}w(\partial_x\ln w)^2,\nn\\
\kl \hat X^2_s\kr&=&\frac14\int_{x,p}\big (\partial_p\psi(x,p)\big)^2=\frac14
\int_{x,p}w(\partial_p\ln w)^2.
\ea
Here we use $\partial_xw=2\psi\partial_x\psi~,~(\partial_xw)^2=4w(\partial_x\psi)^2$ and consider classical probability distributions $w$ for which the r.h.s. of eq. \eqref{113D} is well defined. For products
\be\label{113E}
G_{nm}=\hat X^n_s\hat P^m_s
\ee
one finds $\kl G_{nm}\kr=0$ for odd $n$ or $m$, while for both $n$ and $m$ even we can express $\kl G_{nm}\kr$ in terms of $\partial_x\ln w$ and $\partial_p\ln w$ without an explicit dependence on the sign of $\psi$, similar to eq. \eqref{113D}. We conclude that $\hat X_s$ and $\hat P_s$ are quantum observables for which operators of the type $G$ can be computed in terms of the classical probability distribution $w$. 

\medskip\noindent
{\bf 5. Statistical observables}

The expectation value $\kl P^2_s\kr$ in eq. \eqref{113D} is a measure for the roughness of the probability distribution in position space. Consider the example
\be\label{113F}
w(x)=\int_pw(x,p)=\epsilon^2\cos^2\left(\frac{x}{l}\right)\exp 
\left(-\frac{x^2}{2\Delta^2}\right),
\ee
with $\epsilon$ chosen such that 
\be\label{113G}
W(\epsilon^2,l,\Delta^2)=\int_xw(x)=1.
\ee
Keeping $\epsilon,l$ and $\Delta^2$ as parameters one finds
\be\label{113H}
W(\epsilon^2,l,\Delta^2\pi)=\epsilon^2\left(\frac{\Delta^2\pi}{2}\right)^{1/2}
\left[1+\exp\left(-\frac{2\Delta^2}{l^2}\right)\right],
\ee
and we can easily compute
\ba\label{113I}
\kl \hat X^{2n}_{cl}\kr
&=&W^{-1}\left(2\Delta^4\frac{\partial}{\partial\Delta^2}\right)^nW,\\
\kl \hat X^2_{cl}\kr&=&2\Delta^4(\partial\ln W/\partial\Delta^2)\nn\\
&=&\Delta^2
\left(1-\frac{4\Delta^2}{l^2}
\frac{\exp(-2\Delta^2/l^2)}{1+\exp(-2\Delta^2/l^2)}\right).\nn
\ea
For $l^2/\Delta^2\to 0$ one finds that $\kl X^2_{cl}\kr\to \Delta^2$ becomes independent of $l$ up to exponentially small corrections, while for $l^2/\Delta^2\to\infty$ the dispersion reaches $\Delta^2$ up to power corrections, $\kl X^2_{cl}\kr\to\Delta^2-\Delta^4/l^2$. The situation is qualitatively similar for higher powers $\kl X^{2n}_{cl}\kr$. 

On the other hand, we have
\ba\label{113J}
\kl \hat P^2_s\kr&=&\int_x\big (\partial_x\psi(x)\big)^2,\nn\\
\psi(x)&=&\int_p\psi(x,p)=\epsilon\cos\left(\frac{x}{l}\right)\exp 
\left(-\frac{x^2}{4\Delta^2}\right),
\ea
where we assume here for simplicity a factorized $\psi(x,p)=\epsilon(p)\tilde\psi(x)$ with $\epsilon^2=\int_p\epsilon^2(p)$. We find that
\be\label{113K}
\kl \hat P^2_s\kr=\frac{1}{l^2}
\frac{1}{1+\exp(-2\Delta^2/l^2)}+\frac{1}{4\Delta^2}
\ee
diverges as $\kl \hat P^2_s\kr\approx l^{-2}$ for $l^2/\Delta^2\to 0$. The limit $l\ll\Delta$ clearly demonstrates that $\kl \hat P^2_s\kr$ can detect properties of the probability distribution that are very difficult to be found by measuring powers of $X_{cl}$ and $P_{cl}$. 

The expectation value $\kl \hat P^2_s\kr$ reflects properties of the probability distribution, but it cannot be expressed as a standard classical observable. It is not possible to assign some fixed value $P^2_s(x,p)$ to every classical state $(x,p)$, such that $\kl P^2_s\kr=\int_{xp}w(x,p)P^2_s(x,p)$. Nevertheless, for a suitable class of probability distributions all moments $\kl P^n_s\kr$ are calculable from $w$, such that $P_s$ may be regarded as an observable in this sense. We will call observables of this type ``statistical observables''. They have a status similar to entropy or temperature, which can be assigned to equilibrium ensembles, but have no fixed value in a given microstate. The question if $P^2_s$ can be measured depends on the existence of measurement devices that can be brought into correlation with values of $P^2_s$ in a given range. Formally, we may associate to $P^2_s$ the ``nonlinear observable'' $(\partial_x\ln w)^2/4$, with
\be\label{113L}
\kl P^2_s\kr=\int_{x,p}w\frac{(\partial_x\ln w)^2}{4}.
\ee
We  note, however, that $\kl P^4_s\kr$ differs from $\int_{x,p}w(\partial_x\ln w)^4/16$. 

A very interesting feature of the use of the classical wave function $\psi(x,p)$ and the quantum formalism is the possibility to express statistical observables as $P^2_s$ in the standard way as quantum operators. This allows us to use all the concepts from quantum mechanics, as the spectrum of the observables or the commutation relations and uncertainty relations. In particular, we can make a change of basis such that $P_s$ and $X_s$ are represented as diagonal operators. This defines new states of the system for which $P_s$ and $X_s$ take fixed values. In the quantum formalism, $P_s$ is a perfectly acceptable observable. If it can be measured, the predictions for such measurements are the ones from quantum formalism. 

We will see in sect. \ref{Quantumparticlesfrom} that we can indeed define quantum operators for position and momentum
\ba\label{113M}
\hat X_Q&=&\hat X_{cl}+\frac12\hat X_s,\nn\\
\hat P_Q&=&\hat P_{cl}+\frac12\hat P_s~,~
[\hat X_Q,\hat P_Q]=i,
\ea
whose expectation values can be computed from the classical wave function $\psi(x,p)$ and therefore from the classical probability $w(x,p)$. In particular, the commutator obeys the standard quantum mechanical relation for the position and momentum operators. 

\medskip\noindent
{\bf 6. Classical wave function}

The relation $w=\psi^2$ fixes $\psi$ only up to a sign $s(x,p)=\pm 1$,
\be\label{113N}
\psi(x,p)=s(x,p)w^{1/2}(x,p).
\ee
We will argue next that $s(x,p)$ is severely resticted by properties of continuity and differentiability. Essentially, $\psi(x,p)$ is determined by $w(x,p)$ and does not contain additional information. (Of course, the overall sign of $\psi$ is arbitrary, but it plays no physical role.) If a continuous and arbitrarily often differentiable function $\psi(x,p)$ exists for a given choice of $s(x,p)$, it is clear that a different choice of $s(x,p)$ would not share these properties any longer (except change of overall sign).

We can discuss this issue in terms of the statistical observables $\hat X_s, \hat P_s$ or the ``quantum observables'' $\hat X_{Q},\hat P_Q$. Indeed, the requirement that expectation values of the type $\kl F(\hat X_{Q},\hat P_Q)\kr$ can be computed in terms of $\psi(x,p)$ imposes restrictions on the sign $s(x,p)$. If the expectation value $\kl \hat P^2_s\kr$ or $\kl \hat P^2_Q\kr$ exists, this supposes that expressions of the type
\be\label{113O}
\int_x(\partial_x\psi)^2=\frac14\int_xw(\partial_x\ln w)^2
\ee
are well defined. Thus $\psi$ should be a continuous function of $x$, since a discontinuity would generate a divergent expression on the l.h.s. of eq. \eqref{113O}. Similarly, the existence of
\be\label{113P}
\int_x(\partial^2_x\psi)^2=\frac14\int_xw\big[\partial^2_x\ln w+\frac12(\partial_x\ln w)^2\big]^2
\ee
requires $\partial_x\psi$ to be continuous. Close to a zero of $\psi$ at $x_0$ this allows $\psi=a(x-x_0)$, while a different sign for $x<x_0$, as $\psi=a|x-x_0|$, is not consistent with eq. \eqref{113P}. We can define the expectation values of statistical observables as $\kl P^2_s\kr~,~\kl P^4_s\kr$ in terms of $w(x,p)$. If they are finite, an appropriate choice of $s(x,p)$ exists such that the expressions of the expectation values in terms of operators acting on $\psi(x,p)$ also exist. This restricts $s(x,p)$ by the properties of $w(x,p)$. 

If we assume that arbitrary powers $\kl P^n_s\kr~,~\kl X^n_s\kr$ exist, the sign function $s(x,p)$ is determined by the topology of the zeros of $w(x,p)$. For any $w(x,p)$ without zeros continuity of $\psi$ requires that $s(x,p)$ has to be the same for all points $(x,p)$. Thus a sign flip of $s(x,p)$ can only occur for zeros of $w,w(x_0,p_0)=0$. Assume next that $w(x,p)$ has an ``isolated zero'' at $(x_0,p_0)$ in the sense that $w(x,p)$ is strictly positive in a neighborhood around $(x_0,p_0)$ (excluding the point $(x_0,p_0)$). Then we may draw a circle around $(x_0,p_0)$, arbitrarily close to $(x_0,p_0)$. On this circle $w(x,p)$ is strictly positive, and continuity implies that $s(x,p)$ cannot change sign on the circle. Extending its value to $(x_0,p_0)$ the sign of $s(x,p)$ is the same in a whole neighborhood around $(x_0,p_0)$. This argument extends to zeros on a compact subspace of the phase space ${\mathbbm R}^2$ (with coordinates $x,p)$. The sign must be the same in the region surrounding the subspace where $w$ is strictly positive, and we can formally extend this sign to the subspace where $w$ vanishes.

Typically, $w$ vanishes on the ``boundary of phase space'' for $(x^2+p^2)\to\infty$. We may have a line of zeros from one ``point'' (direction) on the boundary to another, dividing phase space into two pieces. At a given point $(x_0,p_0)$ on the line we may denote by $y$ a coordinate in the direction perpendicular to the line, with $|y|$ the distance from the line. The existence of $\kl\hat P^2_s\kr$ and $\kl \hat X^2_s\kr$ requires that $\sqrt{y}(\partial\sqrt{w}/\partial y)$  vanishes for $y\to 0$, while the existence of higher powers $\kl \hat P^4_s\kr$ etc. implies that the limits for $y\to 0_\pm$ of $\partial\sqrt{w}/\partial y$ exist. (The limits may be different for positive and negative $y$.) For nonzero $\partial\sqrt{w}/\partial y$ the sign $s(x,p)$ must jump at $(x_0,p_0)$, such that $\psi$ has opposite sign on the two different sides of the line, $\lim_{y\to 0}\psi\to ay$. If the line is isolated (no crossing with other lines) it divides ${\mathbbm R}^2$ into two pieces with a jump of $s$ at the division. This type of arguments  can be extended to crossing lines and points on the  line where $a=\lim_{y\to 0}(\partial \psi/\partial y)$ vanishes. For $(\partial\sqrt{w}/\partial y)(y\to 0)=0$ and $(\partial^2\sqrt{w}/\partial y^2(y\to 0)\neq 0$ there is no sign jump, with $\psi\sim by^2$. The problem reduces to the sign of an infinitely often differentiable function $\psi(y)$ with a zero at $y=0$. This type of discussion also extends to $p$-dimensional phase space, where the line is replaced by a $(p-1)$-dimensional hypersurface. 

In summary, for a large class of classical probability distributions $w(x,p)$ the classical wave function $\psi(x,p)$ can be determined in terms of $w(x,p)$. The sign $s(x,p)$ in eq. \eqref{113N} is computable from $w(x,p)$. The expectation values of non-commuting quantum operators, as $\hat X_Q,\hat P_Q$ in eq. \eqref{113M}, or more generally $F(\hat X_Q,\hat P_Q)$, can be computed from the classical probability distribution. When expressed in terms of the classical wave function, they obey the standard laws of quantum mechanics.

\newpage\noindent
{\bf 7. Interference for classical particles}

The particle-wave duality is not a characteristic of quantum particles. Any probabilistic description of particles induces the concept of a particle-wave. These wave aspects are shared by classical particles as well. Even though particles can be viewed as discrete objects, the probability distribution $w(x,p)$ is a continuous function or a field. The time evolution of $w$ specifies the field equation for the wave. We also have introduced the wave function $\psi(x,p)$ for classical particles. It has the properties of a probability amplitude, with $w=|\psi|^2$. The time evolution of $w$ is mapped to a time evolution of $\psi$ - it has the form of a wave equation or Schr\"odinger equation.

If classical particles can be described by waves, with the same interpretation as probability amplitudes as in quantum mechanics, one may wonder what happens with possible interference effects that are usually considered as being characteristic for quantum particles. The generalized Schr\"odinger equation \eqref{89A} is linear in $\psi$ such that the superposition principle holds: if $\psi_1$ and $\psi_2$ are two solutions for the Schr\"odinger equation, also $\psi_1+\psi_2$ is a solution. If the Hamiltonian is given by $H_L$ \eqref{zC}, the time evolution for the probabilities $w=\psi^2$ is given by the linear Liouville equation, such that the superposition principle holds for the probabilities as well. One may ask if in a double slit experiment the probabilities add (no interference) or the amplitudes add (interference). We will see that this depends on the precise initial conditions for the classical probability distribution.

Consider two initial classical wave packets $w_1(x,p,t=0)~,~w_2(x,p,t=0)$, chosen such that for $w_1$ the particle passes through slit 1, and for $w_2$ it passes through slit 2. This is possible since we may have wave packets which correspond to sufficiently focused beams (with appropriate small $\Delta_x,\Delta_p$), such that the time evolution of $w_1(x,p,t)$ leads for all $t$ to a vanishing probability at the location $x_{s2}$ of slit 2, $w_1(x_{s2},p,t)\approx 0$. The initial wave functions corresponding to $w_1$ and $w_2$ are $\psi_1(x,p,t=0)$ and $\psi_2(x,p,t=0)$. If the initial conditions are set by $w(t=0)=w_1(t=0)$ or $w(t=0)=w_2(t=0)$ the particle will indeed ``pass through only one slit'' and no interference pattern will be obtained on a screen behind the slits.

This situation extends to an initial probability distribution
\be\label{W1}
w(t=0)=aw_1(t=0)+(1-a)w_2(t=0).
\ee
Due to the superposition principle for $w$ the probability for finding hits on the screen at the location which correspond to particles which have passed through slit 1 amounts to $a$, and to $1-a$ for the location corresponding to slit 2. The initial wave function for this setting is 
\be\label{W2}
\psi(t=0)=\big (aw_1(t=0)+(1-a)w_2(t=0)\big)^{1/2}.
\ee

One may, however, prepare also a different initial probability distribution,
\ba\label{W3}
\psi(t=0)&=&a_1\psi_1(t=0)+a_2\psi_2(t=0),\nn\\
w(t=0)&=&a_1^2w_1(t=0)+a^2_2w_2(t=0)\nn\\
&&+2a_1a_2\psi_1(t=0)\psi_2(t=0),
\ea
with $a_2$ determined by $a_1$ and the normalization of $\psi$ or $w$. Now the superposition principle for $\psi$ implies $\psi(t)=a_1\psi_1(t)+a_2\psi_2(t)$ or
\ba\label{W4}
w(x,p,t)&=&a^2_1w_1(x,p,t)+a^2_2w_2(x,p,t)\nn\\
&&+2a_1a_2\psi_1(x,p,t)\psi_2(x,p,t).
\ea
We recognize an interference term $\sim a_1a_2$ which leads, in principle, to an interference pattern on the screen. The interference can be positive or negative. It becomes an important effect if $a_1$ and $a_2$ are of similar magnitude and $\psi_1$ and $\psi_2$ have a substantial overlap for some positions on the screen. The details of the interference differ, however, from the quantum particle. They also depend on the question which location observable is appropriate for measuring the location of the particle on the screen, e.g. $\hat X_{cl}$ or $\hat X_Q$.

\section{Quantum and classical particles in a potential}
\label{inapotential}
In this section we compare the propagation of a quantum particle and a classical particle in a given potential $V(x)$. We use for both a probabilistic description with a distribution of initial values for location and momentum at time $t_0=0$. We also employ for both the quantum mechanical formalism, keeping in mind that the latter can ultimately be obtained from a classical statistical ensemble. Using the same formalism constitutes an appropriate framework for a discussion of the differences between quantum and classical particles. We have seen already that the expectation values of position and momentum obey the same evolution equation \eqref{zE}. Possible differences must then be connected to a different evolution of the probability distribution, or differences in the allowed initial conditions. 

\medskip\noindent
{\bf 1. One particle phase space distribution}

As a starting point for the classical particle we consider the probability in phase space $w(x,p)$, from which observables involving arbitrary functions of position and momentum can be computed
\be\label{F1}
\kl F(x,p)\kr=\int_x\int_pF(x,p)w(x,p).
\ee
(Here we use in $d$ dimensions the shorthands $\int_x=\int d^dx$, $\int_p=(2\pi)^{-d}\int d^dp$. We keep our discussion one dimensional, but many formulae can be extended to arbitrary $d$ if suitable scalar products of vectors are used when appropriate.) 

We will compare $w(x,p)$ with the Wigner representation $\rho_w(z,q)$ of the density matrix for a quantum particle, as given for the free particle by eq. \eqref{ZQ}. From the density matrix $\rho(x,y)$ in position space the Wigner representation obtains as
\be\label{F2}
\rho_w(z,q)=\int d(x-y)e^{-iq(x-y)}\rho(x,y),
\ee
where $z=(x+y)/2$ is kept fixed for the integration over the ``relative coordinate'' $(x-y)$. We next recall in simple terms that appropriate correlation functions for a quantum particle obey eq. \eqref{F1} if $w(x,p)$ is replaced by $\rho_w(x,p)$ \cite{Moyal}.

The expectation value of an observable which is a function $F_x(x)$ of the position obeys
\ba\label{F3}
\kl F_x(x)\kr&=&\text{tr}\big(\rho F_x(\hat X)\big)=\int_x\rho(x,x)F_x(x)\nn\\
&=&\int_x\int_pF_x(x)\rho_w(x,p).
\ea
Here we use the inverse Fourier transform in order to express $\rho(x,y)$ in terms of the Wigner representation $\rho(z,q)$
\be\label{F4}
\rho(x,y)=\int_pe^{ip(x-y)}\rho_w\left({\frac{x+y}{2},p}\right),
\ee
with
\be\label{F5}
\rho(x,x)=\int_p\rho_w(x,p),
\ee
and $\hat X$ is the usual position operator in quantum mechanics. Similarly, we find for a function $F_p(p)$ of momentum
\ba\label{F6}
\kl F_p(p)\kr&=&\text{tr}\big(\rho F_p(\hat P)\big)=\lim_{y\to x}\int_xF_p\left(-i\frac{\partial}{\partial x}\right)\rho(x,y)\nn\\
&=&\int_x\int_pF_p\left(p-\frac i2\frac{\partial}{\partial x}\right)\rho_w(x,p)\nn\\
&=&\int_x\int_pF_p(p)\rho_w(x,p),
\ea
with $\hat P$ the quantum mechanical momentum operator. For the last equation \eqref{F6} we assume the absence of boundary terms, as appropriate if $\rho_w(x,p)$ has support only in a local region of space. Due to linearity, eqs. \eqref{F3}, \eqref{F6} extend to all observables of the form $F(x,p)=F_x(x)+F_p(p)$, and we find the expression analogous to eq. \eqref{F1} 
\be\label{F7}
\kl F(x,p)\kr=\int_x\int_pF(x,p)\rho_w(x,p).
\ee

For observables involving both powers of $\hat X$ and $\hat P$ the order of the operators matters. One finds
\ba\label{F8}
\text{tr}(\hat X\hat P\rho)&=&\int_x\int_px
\left(p-\frac i2\partial_x\right)\rho_w(x,p)\nn\\
&=&\int_x\int_p\left(xp+\frac i2\right)\rho_w(x,p),
\ea
while 
\be\label{111A}
\text{tr}(\hat P\hat X\hat \rho)=\int_x\int_p\left(xp-\frac i2\right)\rho_w(x,p).
\ee
In particular, the correlation function for subsequent measurements of $x$ and $p$ \cite{CWAA},
\be\label{F10}
\kl XP\kr_m=\frac 12\text{tr}\big(\{\hat X,\hat P\}\rho\big)=\int_x\int_pxp\rho_w(x,p)
\ee
is given by the same expression as the correlation function for a classical particle if $\rho_w$ replaces $w$ in eq. \eqref{F1}. The expectation values of arbitrary sequences of operators $\hat X$ and $\hat P$ can be represented as 
\be\label{F11}
\kl S[\hat X,\hat P]\kr=\int_x\int_pS[X_Q,P_Q]\rho_w(x,p),
\ee
where 
\be\label{F12}
X_Q=x+\frac i2\frac{\partial}{\partial p}~,~P_Q=p-\frac i2\frac{\partial}{\partial x},
\ee
and the order in the sequence is the same on both sides of eq. \eqref{F11}. As it should be, the commutation relation is transferred from the operators $\hat X$ and $\hat P$ to $X_Q$ and $P_Q$,
\be\label{F13}
[X_Q,P_Q]=i.
\ee
The r.h.s. of eq. \eqref{F11} assumes that partial integrations can be performed, i.e. that boundary terms where derivative operators $\partial_p$ or $\partial_x$ stand on the first place in the sequence (``on the left'') vanish. 

In particular, we may consider the totally symmetrized functions of operators $F_s(\hat X,\hat P)$, where we assume that an expansion of $F(x,p)$ in powers of $x$ and $p$ is possible. It is defined by associating to each factor the totally symmetrized combination of all possible ordered sequences, as
\ba\label{F14}
(\hat X^2\hat P^2)_s&=&\frac16(\hat X^2\hat P^2+\hat P^2\hat X^2+\hat X\hat P^2\hat X+
\hat P\hat X^2\hat P\nn\\
&&+\hat X\hat P\hat X\hat P+\hat P\hat X\hat P\hat X).
\ea
Employing methods similar to \cite{CWNEQ} one finds the simple expression
\be\label{F15}
\kl F_s(\hat X,\hat P)\kr=\int_x \int_pF(x,p)\rho_w(x,p).
\ee
This generalizes eq. \eqref{F7}. Comparison with eq. \eqref{F1} demonstrates the close formal correspondence between classical and quantum particles. We may define the one-particle phase  space distribution $f_1(x,p)$ as $f_1(x,p)=w(x,p)$ for classical particles and $f_1(x,p)=\rho_w(x,p)$ for quantum particles. The function $f_1(x,p)$ generates all expectation values of totally symmetrized products of observables according to eq. \eqref{F1} or \eqref{F15}.

\medskip\noindent
{\bf 2. Differences between quantum and classical 

\hspace{0.2cm}particles}

For a given distribution $f_1(x,p)$ the only difference between a quantum particle and a classical particle concerns the ordering of sequences of measurements of the observable $X$ or $P$. Only the totally symmetrized orderings coincide. For the classical particle the ordering does not matter for the classical correlation functions since $\hat X_{cl}$ and $\hat P_{cl}$ commute. This does not hold for a quantum particle, where the measurement correlations depend on the ordering of sequences of measurements \cite{CWAA}. For the quantum particle one finds for the measurement correlation for sequences of measurements in different orders (the first measurement corresponds to the observable on the right of the sequence)
\ba\label{F16}
\kl PXPX\kr_m&+&\kl XPXP\kr_m-\kl X^2P^2\kr_m-\kl P^2X^2\kr_m\nn\\
&=&\frac12\text{tr}\Big\{\rho\big([\hat X,\hat P]\big)^2\Big\}=-\frac12.
\ea
It is possible, however, to define also for the classical particle a different set of correlation functions. We show in appendix B such an alternative definition which shares all properties of the quantum correlations. 

A second important difference between a quantum particle and a classical particle concerns the time evolution of the one-particle phase space distribution. For the classical particle this is given by the Liouville equation
\be\label{119A}
\partial_tw(x,p)=\left\{-\frac pm\partial_x+\frac{\partial V(x)}{\partial x}\partial_p\right\}
w(x,p).
\ee
In contrast, for a quantum particle one finds
\ba\label{119B}
&&\partial_t\rho_w(x,p)=
\left\{-\frac pm\partial_x+iV\left(x-\frac i2\partial_p\right)\right.\nn\\
&&\qquad~~ \qquad \left.-iV\left(x+\frac i2\partial_p\right)\right\}\rho_w(x,p)\\
&&=\left\{-\frac pm\partial_x+2 V(x)\sin\left(\frac 12
\stackrel{\leftarrow}{\partial_x} \stackrel{\rightarrow}{\partial_p}\right)\right\}
\rho_w(x,p).\nn
\ea
For this purpose we assume the von-Neumann equation for the time evolution of the density matrix for the quantum particle
\be\label{119C}
\partial_t\rho=-i[H_Q,\rho],
\ee
with
\be\label{119D}
H_Q=\frac{\hat P^2}{2m}+V(\hat X).
\ee
This yields, for $z=(x+y)/2$, the relation
\ba\label{119E}
\partial_t\rho(x,y)&=&\int_p\partial_t\rho_w(z,p)e^{ip(x-y)}\\
&=&\int_p\left\{-iV(x)+iV(y)-
\frac pm\partial_z\right\}
\rho_w(z,p)e^{ip(x-y)},\nn
\ea
from which we infer eq. \eqref{119B} by partial integration. 

For a formal Taylor expansion of $V$ in eq. \eqref{119B} in $i\partial_p$ one finds
\be\label{119F}
\partial_t\rho_w(x,p)=
\left\{-\frac pm\partial_x+\frac{\partial V}{\partial x}\partial_p-\frac{1}{24} 
\frac{\partial^3 V}{\partial x^3}\partial^3_p+\dots\right\}
\rho_w(x,p)
\ee
The first term agrees with the classical particle \eqref{119A}. In particular, for a free particle or for a harmonic potential $V\sim x^2$ there is no difference between the time evolution of $f_1(x,p)$ for a classical or a quantum particle. For a given initial value of $f_1(x,p)$ the time evolution of all totally symmetrized correlation functions is then the same for classical and quantum particles.

In summary, the time evolution of the one particle distribution for classical and quantum particles has a similar structure. For the classical particle it is given by the Liouville equation
\be\label{160AA}
\partial_t w=-iH_L w,
\ee
with $H_L$ given by eq. \eqref{zC}. The quantum particle is characterized by a different evolution operator
\be\label{160BB}
\partial_t\rho_w=-i H_W\rho_w,
\ee
with $H_W$ obeying
\be\label{GP2}
H_W=-i\frac pm\partial_x+V\left(x+\frac i2\partial_p\right)-V
\left(x-\frac i2\partial_p\right).
\ee

Finally, a third difference concerns the allowed phase space distributions $f_1(x,p)$. In both cases the normalization is the same $\int_x\int_pf_1(x,p)=1$. For a classical particle $w$ is a probability density in phase space and therefore $f_1(x,p)\geq 0$ must hold everywhere in phase space. For a quantum particle $\rho_w(x,p)$ is not anymore a probability density. Only the diagonal elements of the density matrix must be positive, i.e.
\be\label{119G}
\rho(x,x)=\int_p\rho_w(x,p)\geq 0~,~
\rho(p,p)=\int_x\rho_w(x,p)\geq 0.
\ee
This does not exclude $\rho_w(x,p)$ to be negative in certain regions in phase space.

On the other hand, not every classical $f_1 (x,p)$ corresponds to a quantum mechanical density matrix. While the positivity and normalization of the diagonal elements $\rho (x,x)$ or $\rho (p,p)$ (eq. \eqref{119G} ) is obeyed automatically, this is not sufficient to guarantee the positivity of the density matrix. For example, a quantum mechanical density matrix has to obey the constraint
\be\label{Y1}
\text{tr}\rho^2\leq 1~,~\int_{x,y}\rho(x,y)\rho(y,x)=\int_{x,y}|\rho(x,y)|^2\leq 1.
\ee
In terms of the Wigner representation this condition reads
\be\label{Y2}
\text{tr}\rho^2=\int_{x,y}|\rho(x,y)|^2=\int_{z,p}|\rho_w(z,p)|^2\leq 1.
\ee
Thus an ensemble of free classical particles can mimic the one particle phase space distribution for quantum particles with $\rho_w=w$ only if the ``purity constraint''
\be\label{Y3}
\int_{z,p}w^2(z,p)\leq 1
\ee
is obeyed. In this case one may expect the quantum mechanical uncertainty relation for the location and momentum observables to hold. For a Gaussian initial wave packet \eqref{zF} the purity constraint reads $(w^2=|\psi|^4)$
\ba\label{Y4}
&&\int_{z,p}(\Delta_x\Delta_p)^{-2}\exp
\left\{-\frac{(x-\bar x)^2}{\Delta^2_x}\right\}
\exp\left\{-\frac{(p-\bar p)^2}{\Delta^2_p}\right\}\nn\\
&&\hspace{1.0cm}=\frac{1}{2\Delta_x\Delta_p}\leq 1.
\ea
This is precisely the uncertainty relation $\Delta_x\Delta_p\geq 1/2$, which is not automatically obeyed by all classical $w(x,p)$.

Indeed, a quantum mechanical density matrix has to obey a positivity constraint, namely that all eigenvalues of the matrix $\rho(x,y)$ must be positive or zero. In turn, this imposes a constraint on the classical probability densities which reproduce a phase space description for a quantum particle: the hermitean matrix
\be\label{Y5}
\tilde w(x,y)=\int_p e^{ip(x-y)}w\left(\frac{x+y}{2},p\right)
\ee
should only have positive or zero eigenvalues. We recall the properties
\ba\label{Y6}
\tilde w(x,x)&=&\int_p \tilde w(z=x,p) \geq 0,\nn\\
\tr \tilde w&=&\int_x \tilde w(x,x)= 1.
\ea 
Provided the classical probability distribution $w(z,p)$ is square integrable and therefore obeys the purity constraint \eqref{Y3}, also the condition $\tr w^2 \leq 1$ holds. If we consider the continuous function $\tilde w(x,y)$ as the limit of a sequence with a finite number of degrees of freedom $M$, the purity constraint \eqref{Y3} follows from the positivity of $\tilde w(x,y)$ by a suitable unitary transformation, $\tilde w(x,y) \to p(x) \delta_{x,y}$. The conditions $p(x)\geq0,\sum_x p(x)=1$ imply $\sum_x p^2(x)\leq 1$ and therefore 
$\tr~ \tilde w^2\leq 1$. Inversely, the constraint \eqref{Y3} and the properties \eqref{Y6} are not sufficient in order to guarantee the positivity of $\rho(x,y)$, the latter being the strongest condition for $f_1(z,p)$ to describe a quantum particle. 

We conclude that the sets of allowed phase space distributions for quantum and classical particles have an overlap, namely whenever $f_1(z,p) \geq 0$ and the Fourier transform $f_1(x,y)$, given by eq. \eqref{F4} or \eqref{Y5}, is a positive matrix. Certain states are allowed only for classical particles ($f_1(z,p)\geq0, f_1(x,y)$ has negative eigenvalues) or only for quantum particles ($f_1(z,p)$ becomes negative in certain regions of phase space, $f_1(x,y)$ is a positive matrix). Certain functions $f_1$, where neither $f_1(x,y)$ is a positive matrix nor $f_1(z,p)$ is positive semidefinite, can be realized neither by a quantum nor by a classical particle. It would be interesting to know if generalizations of the particle concept beyond classical and quantum particles could realize such functions.

Despite the close formal similarity between quantum and classical particles we summarize that differences arise on three levels. (i) The different status of $w$ and $\rho_w$ implies different restrictions on the allowed states, in particular the allowed initial values. While $w$ is a diagonal density matrix, $\rho_w$ is the Fourier transform of a density matrix with respect to only the relative coordinate $x-y$ and therefore not a density matrix by itself. (ii) For unharmonic potentials the different Hamiltonians generating the time evolution of classical and quantum particles are responsible for a different time evolution of the phase-space density $f_1(x,p)$. (iii) The non-commutativity of the operators for a quantum particle entails a dependence of correlations on the sequence of measurements.

\section{Quantum particles from classical probabilities}
\label{Quantumparticlesfrom}

The conceptual unification of quantum and classical particles suggests that both could be based on a common description in terms of a classical probability distribution in phase space. This possibility can also be regarded in the light of the limiting processes discussed in sect. \ref{particlemotion}. The quantum particle requires in a sense less information than the classical particle. In eqs. \eqref{43A}, \eqref{43B} the quantum observables use only a subspace. In a separate paper we will show in detail that such a description of a quantum particle moving in an arbitrary potential is indeed possible. 

In this short section we only sketch how all the properties of a quantum particle can be described in terms of classical probabilities in phase space. For this purpose we use the classical wave function $\psi(x,p)$, which contains the same information as the classical probability distribution $w(x,p)=\psi^2(x,p)$. As compared to the classical particle, the quantum particle needs two modifications: (i) the use of quantum observables instead of classical observables, and (ii) a different time evolution of the classical probability density in phase space.

Quantum observables $\hat X_Q$ and $\hat P_Q$ are defined by the operators \eqref{F12}. They obey the usual commutation relations for the quantum operators for position and momentum. The expectation value for an arbitrary sequence of such observables can be computed from the classical probability distribution by the usual quantum rule
\be\label{J1}
\kl F(X_Q,P_Q)\kr=\int_{x,p}\psi(x,p)F(\hat X_Q,\hat P_Q)\psi(x,p).
\ee
We recall that the quantum observables involve the statistical observables $X_s$ and $P_s$. If the ensemble is characterized by the classical probability distribution for states associated to the phase-space points $(x,p)$, the quantum position and momentum do therefore not correspond to classical observables with a fixed value in every state. (If we want to realize $\hat X_Q,\hat P_Q$ as classical observables we have to implement them in an ensemble with a larger set of states, as discussed in sect. \ref{quantumandclassical}.)

For a quantum particle, the time evolution of the classical wave function (and the associated classical probability distribution in phase space) is given by
\be\label{J2}
i\partial_t\psi(x,p)=H_W\psi(x,p),
\ee
where we have replaced the Hamiltonian $H_L$ by a modified Hamiltonian $H_W$ given by eq. \eqref{GP2}. Eq. \eqref{J2} should be regarded as the new fundamental equation describing the dynamics of particles. In consequence, the classical probability density $w(x,p)$ obeys a new non-linear time evolution equation instead of the Liouville equation. We can express the Wigner representation of the quantum mechanical density matrix in terms of the classical wave function by 
\ba\label{AA1}
&&\bar\rho_w(x,p)=\\
&&\int_{r,r',s,s'}
\psi(x+\frac r2,p+s)\psi(x+\frac{r'}{2},p+s')\cos (s'r-sr').\nn
\ea  
It is straightforward to verify that with eq. \eqref{J2} the time evolution of $\bar\rho_w$ obeys the usual time evolution for the density matrix of a quantum particle, with Hamiltonian $H_Q$ given by eq. \eqref{119D}. In other words, the time evolution of $\bar\rho_w$ as defined by eq. \eqref{AA1} obeys eq. \eqref{119B}. We leave  the proof that $\bar\rho_w(x,p)$, as defined by eq. \eqref{AA1}, obeys the necessary  positivity conditions for the associated density matrix to a separate paper. We will  also show that the one particle distribution function for the quantum particle can be  understood as a coarse graining of the classical probability distribution in phase space. 

In summary, the expectation values of the quantum observables $X_Q$ and $P_Q$ and all their correlation functions obey all the relations for a quantum particle in a potential, including their time evolution. Starting at some initial time $t_0$ with a classical probability distribution which corresponds to a given $\rho_w(x,p)$, all quantum laws for a quantum  particle in a potential are obeyed for all times, including characteristic phenomena as interference and tunneling. These correlation functions are the only thing measurable in this system - demonstrating that quantum mechanics can be described in terms of a classical probability distribution in phase space.

\section{Conclusions and discussion}
\label{conclusionsand}
We have realized the description of a quantum particle in a setting of a classical statistical ensemble with infinitely many degrees of freedom. While the classical statistical  ensemble contains information about the particle and its environment, the typical quantum mechanical features emerge if we concentrate on the statistical description of the subsystem for the particle. For the subsystem we deal with ``incomplete statistics'' for which joint probabilities can no longer be used for the prediction of outcomes of measurements of two observables. This is the origin for the representation of such observables by non-commuting operators in the associated quantum formalism. We have derived the quantum mechanical operator representation of the relevant position and momentum operators for a quantum particle explicitly, starting from a simple ``one-bit particle'', generalizing to ``$Q$-bit particles'', and finally taking the continuum limit $Q\to\infty$ in order to obtain position and momentum operators with a continuous spectrum. Heisenberg's law for the commutator of position and momentum operators and the associated quantum mechanical uncertainty principle follow.

In the course of this construction we have seen that alternative selections of observables, which lead in the continuum limit to commuting position and momentum operators, are also possible. The same system of states and observables (the same Hilbert space) can therefore describe both quantum and classical particles. Quantum and classical particles are characterized, however, by different Hamiltonians. The appropriate Hamiltonian for real particles in nature can be (and has been) tested experimentally. For example, the double split experiment shows the interference pattern characteristic for a quantum particle. In contrast, classical particles allow for states with arbitrarily sharp values of the location and momentum observables. If the probability distribution is sufficiently sharp, only particles with trajectories passing by one of the splits will pass, even if the separation of the two splits is of the same order as the quantum mechanical wave length of the particles. This has clearly been falsified by experiment.

The description of both quantum and classical particles by a common formalism gives rise to the interesting question why nature prefers quantum particles. Interesting hints may come from an investigation of stability properties of the quantum and classical wave functions that we have not yet investigated. We also note that the quantum particle is in a sense ``minimal'' since it uses less commuting operators as the classical particle. Furthermore, for the quantum particle the energy and the momentum operators are directly related to translations in time and space, in contrast to the classical particle. 

As a further aspect of the conceptual unification of quantum and classical particles we have developed a quantum formalism for classical particles. It is based on the classical wave function in phase space $\psi(x,p)$, which equals the square root of the probability density up to a sign. Within this formalism the basic equation describing the dynamics of particles is given by a type of Schr\"odinger equation for $\psi$. For classical particles this replaces Newton's equations for trajectories - the latter only emerge as a particular case for infinitely sharp probability distributions. The use of the classical wave function allows for a simple description of new types of statistical observables that measure properties as the ``roughness'' of the probability distribution in position and momentum space. The concept of particle-wave duality also applies to classical particles. In particular, we have discussed interference effects for classical particles. 

Finally, our formulation of the time evolution of a classical probability distribution $w(x,p)$ in terms of a quantum mechanical wave function $\psi(x,p)$ may be used as a formal tool for the study of classical probability distributions. For example, we can immediately give a simple proof that distributions of non-interacting particles, moving in an arbitrary potential $V(x)$, cannot reach the thermal equilibrium distribution as time goes to infinity. The unitarity of the time evolution of $\psi$ leads to obstructions. We can construct an infinity of classical observables $A(x,p)$ which commute with $H$ and are therefore conserved. Such observables obey the differential equation
\be\label{CD}
\frac pm\partial_xA(x,p)=\frac{\partial V}{\partial x}\partial_pA(x,p).
\ee
A simple family of solutions of eq. \eqref{CD} is 
\be\label{CE}
A(x,p)=f(E)~,~E=V(x)+\frac{p^2}{2m},
\ee
with an arbitrary function $f(E)$. 
Thus arbitrary functions of the energy are conserved \cite{AW1}. An infinity of conserved quantities, which have to keep their initial values, contradicts the assumption that for large $t$ a probability distribution of thermal equilibrium is reached, where all expectation values of observables can only depend on two parameters, namely the temperature and the chemical potential. We also find an infinity of static probability distributions in phase space. The corresponding static wave functions obey
\be\label{CEa}
\frac pm \partial_x \psi(x,p)=\frac{\partial V}{\partial x}\partial_p\psi(x,p).
\ee
Again, all functions of the form $\psi(E)$ are static solutions. 

The observation that quantum mechanical features can obstruct the approach of classical probability distributions to thermal equilibrium distributions has been made earlier for classical field theories with interactions \cite{4}. Similar obstructions for interacting particles exist and we believe that it is likely that obstructions beyond the conserved energy distributions are present. This does not contradict the fact that many correlation functions approach the thermal limit for $t\to\infty$, as demonstrated by numerical solutions in simple systems \cite{AW1}. Of course, such obstructions can also be seen in the standard classical formalism. However, the quantum formalism makes them particularly transparent and perhaps more easy to approach. This could also hold for other features of classical probability distributions.

Classical statistics and quantum mechanics are two sides of the same medal, rather than mutually exclusive concepts. This opens the potential for cross-fertilization between the two formalisms, since they both describe the same physical reality. We do not know yet how far the practical use of this unification of concepts will reach.

\LARGE

\section*{APPENDIX A: EXTENDED HAMILTONIAN FOR CLASSICAL PARTICLES}
\renewcommand{\theequation}{A.\arabic{equation}}
\setcounter{equation}{0}

\normalsize
The Hamiltonian $H_L$ \eqref{zC} commutes with the ``classical Hamiltonian''
\be\label{113A}
H\C=\frac{p^2}{2m}+V(x)~,~[H_L,H\C]=0.
\ee
The expectation value of $H\C$ is therefore conserved and can be associated with the classical energy. As we have argued, $H\C$ has no influence on the evolution of $w(x,p)$. We can therefore define an extended Hamiltonian for the classical particle
\be\label{113B}
H=H\C+H_L=\frac{p^2}{2m}+V(x)-i\frac{p}{m}\partial_x+iV'(x)\partial_p.
\ee
The piece $H\C$ does not influence the time evolution of the expectation value of any ``diagonal observable'' $A(x,p)$, which remains purely dictated by $H_L$. However, the use of $H$ brings the quantum mechanical description of the classical particle even closer to the usual setting. For example, a stationary state with fixed energy $E$ obeys
\be\label{113C}
H\C\psi=E\psi~,~H_L\psi=0~,~H\psi=E\psi.
\ee
The stationary wave function takes the form
\ba\label{113DA}
\psi&=&w^{1/2}(\epsilon)\exp (-i\epsilon t),\nn\\
\epsilon(x,p)&=&\frac{p^2}{2m}+V(x),
\ea
with
\be\label{113E}
\epsilon w^{1/2}(\epsilon)=E w^{1/2}(\epsilon)~,~\int_x\int_p w(\epsilon)=1.
\ee
For a fixed energy $w^{1/2}(\epsilon)$ is a $\delta$-type distribution with $w^{1/2}(\epsilon\neq E)=0$. We note that the wave function \eqref{113D} remains an eigenvalue of $H_L$ with eigenvalue zero also for more general $w^{1/2}(\epsilon)$. This implies $H\psi=H\C\psi$ for such states, and $H$ plays the role of the classical energy.

Even more generally, the Schr\"odinger equation $\partial_t\psi=-i(H\C+ H_L)\psi$ has always a solution
\be\label{113F}
\psi(x,p;t)=w^{1/2}(x,p;t)\exp \left\{i\left(\frac{p^2}{2m}+V(x)\right)t\right\},
\ee
with
\be\label{113G}
\partial_t w^{1/2}=-iH_Lw^{1/2}=-\hat L w^{1/2}.
\ee
For this solution one has $\kl H_L\kr=0$ such that the average classical energy obeys
\be\label{113H}
\kl E\kr=\int_{x,p}\psi^*H\C\psi=\int_{x,p}\psi^* H\psi.
\ee
Eq. \eqref{113F} is not the most general solution of the Schr\"odinger equation. Nevertheless, eq. \eqref{113G} continues to hold even for the most general solution. This can be recovered formally for the extended Hamiltonian \eqref{113B}
\ba\label{113I}
\partial_t w&=&\partial_t(\psi^*\psi)=\psi^*\partial_t\psi+(\partial_t\psi^*)\psi\nn\\
&=&-i\psi^*\big[(H\C+H_L)\psi\big]+i\big[(H^*\C+H^*_L)\psi^*\big]\psi\nn\\
&=&-i\big(\psi^*[H\C\psi]-[H\C\psi^*]\psi\big)-i\big(\psi^*[H_L\psi]+[H_L\psi^*]\psi\big)\nn\\
&=&-\hat L w,
\ea
where we use $H^*\C=H\C~,~H^*_L=-H_L$ and the fact that $iH_L$ is a real first order differential operator, while $\psi^*[H\C\psi]=[H\C\psi^*]\psi=\big[p^2/(2m)+V(x)\big]w$. With
\be\label{113J}
\partial_t w=2 w^{1/2}\partial_t w^{1/2}=-\hat L w=-2 w^{1/2}\hat L w^{1/2}
\ee
we verify eq. \eqref{113G}. 

The most general solution of the Schr\"odinger equation can then be written in terms of 
$\psi(x,p;t)$ in eq. \eqref{113F} as
\be\label{113K}
\tilde\psi(x,p;t)=\psi(x,p;t)e^{i\tilde \alpha(x,p;t)}.
\ee
With
\ba\label{113L}
\partial_t\tilde\psi&=&\partial_t\psi e^{i\tilde\alpha}+i\partial_t\tilde\alpha\tilde\psi\\
&=&-i(H\C+H_L)\tilde\psi\nn\\
&=&-i\big[(H\C+H_L)\psi\big]e^{i\tilde\alpha}
-i\psi[H_Le^{i\tilde\alpha}\big]\nn
\ea
we infer that $\tilde\alpha$ evolves according to
\be\label{113MA}
\partial_t\tilde\alpha=-iH_L\tilde\alpha=-\hat L\tilde \alpha.
\ee
According to eq. \eqref{Z4} one finds
\be\label{113NA}
\kl H_L\kr=\int_{xp}w(\hat L\tilde \alpha)=-\int_{xp}w\partial_t\tilde\alpha.
\ee
Since the phase $\tilde\alpha$ is not observable by the use of ``classical observables'' $A(x,p)$ we may use $\tilde \alpha=0$, which remains conserved by the time evolution. With the ``initial condition'' $\tilde \alpha(x,p;t_0)=0$ the solution \eqref{113F}, \eqref{113G} holds for all $t$. 

We recall, however, that there is no need for an extension of the classical Hamiltonian according to eq. \eqref{113B}. In the main text we will continue to describe the time evolution for the classical particle by the Hamiltonian $H_L$. A real wave function $\psi=\pm \sqrt{w}$, with $\alpha\equiv 0,\pi$ in eq. \eqref{Z2}, is then a natural setting. In this case the energy is an independent observable $E=H\C$, not identical with the generator of the time evolution $H_L$. It commutes with $H_L$ and is therefore conserved.

\LARGE
\section*{APPENDIX B: QUANTUM CORRELATIONS FOR CLASSICAL OBSERVABLES}
\renewcommand{\theequation}{B.\arabic{equation}}
\setcounter{equation}{0}
\label{Quantumcorrelationsfor}

\normalsize
Let us consider a statistical ensemble for classical particles for which the initial phase space distribution is chosen such that $\tilde w(x,y)$ in eq. \eqref{Y5} is a positive matrix. For a free particle or a particle in a harmonic potential the phase space distribution $f_1(z,p)$ is then identical to the one of a quantum particle at all later times. The expectation values of all classical correlation functions are equal to the symmetrized correlation functions for the quantum particle at all times. There seems to be no experimental way to distinguish between a classical and a quantum particle in this case, suggesting that the classical and quantum particle are actually identical.

Nevertheless, due to the non-commutativity of the position and momentum operators in quantum mechanics, one can also define correlation functions for an order of operators differing from the totally symmetrized ordering \eqref{F14}. While certain sequences of anticommutators yield the symmetrized product, as 
\be\label{C1}
\frac18\Big\{\big\{\{\hat P,\hat X\},\hat P\big\},\hat X\Big\}=
\frac14\big\{\{\hat P^2,\hat X\},\hat X\big\}=
(\hat X^2\hat P^2)_s,
\ee
(and similarly for $\hat X$ and $\hat P$ exchanged), other orderings differ, as 
\be\label{C2}
\frac12\{\hat P^2,\hat X^2\}=(\hat X^2\hat P^2)_s-\frac12.
\ee
For a quantum system, the choice of the appropriate correlation function depends on the specific question. There is no unique definition of a correlation function for two powers of $X$ and $P$.

The same holds for a classical particle as well. For the special case where the classical probability distribution equals the Wigner transform of an appropriate quantum mechanical density matrix, $w(x,p)=\tilde \rho_w(x,p)$, it is straightforward to construct correlation functions for the classical observables which differ from the classical or pointwise correlation function, in complete analogy to the quantum particle. In order to see this, we assume that $\tilde w(x,y)$ in eq. \eqref{Y5} is a positive matrix and formulate for this case the quantum mechanical concepts and laws directly in terms of the classical probability distribution $w(z,p)$ or its partial Fourier transform $\tilde w(x,y)$. In particular, a pure state matrix $\tilde \rho_w=w$ is realized by a classical probability distribution with the ``folding property'' $\tilde w^2=\tilde w$ or
\be\label{C3}
\int_v\tilde w(x,v)\tilde w(v,y)=\tilde w(x,y).
\ee
It can be described in the usual way by a complex quantum mechanical wave function $\varphi_Q(x)$, with 
\be\label{C4}
\tilde w(x,y)=\varphi_Q(x)\varphi^*_Q(y).
\ee
(We use there the index $Q$ in order to distinguish $\varphi_Q(x)$ from the ``classical wave function'' $\psi_C(z,p)$ which obeys $|\psi_C(z,p)|^2=w(z,p)$.) 

We recall that we can associate to an arbitrary classical observable $A(z,p)$ a quantum mechanical operator $\hat A$
\be\label{C5}
A(z,p)\to \hat A =A_s(\hat X,\hat P),
\ee
where $A_s$ denotes the totally symmetrized product of the operators in a Taylor expansion (cf. eq. \eqref{F14}). The operators $\hat X, \hat P$ act on $\tilde w(x,y)$ or $\varphi_Q(x)$ as
\ba\label{C6}
\hat X\tilde w(x,y)&=&x\tilde w(x,y)~,~\hat P \tilde w(x,y)=-i\partial_x 
\tilde w(x,y),\nn\\
\hat X\varphi_Q(x)&=&x\varphi_Q(x)~,~\hat P\varphi_Q(x)=-i\partial_x\varphi_Q(x).
\ea
We have already shown the quantum laws for expectation values
\ba\label{C7}
\kl A(z,p)\kr&=&\int_{z,p} A(z,p)w(z,p)=\tr(\hat A w)\\
&=&\int_x\hat A \tilde w(x,y)_{|y=x}=
\int_x\varphi^*_Q(x)\hat A\varphi_Q(x),\nn
\ea
where the last relation holds for pure states. In contrast to the classical observables $A(z,p)$ the corresponding quantum operators $\hat A$ do not commute. We can use them in order to derive Heisenberg's uncertainty relation for all classical probability distributions that can be described by a positive matrix $\tilde w(x,y)$ such that they obey the purity constraint \eqref{Y3}. This demonstrates in a simple way how the quantum formalism with non-commuting operators arises from a classical ensemble describing a probability distribution of classical particles. Correlation functions different from the classical correlation function can now be defined by different orderings of the operators $\hat X$ and $\hat P$. 

We close this appendix by the remark that a judgment if a given classical probability distribution $w(x,p)$ can correspond to a quantum state may be often rather involved, since the positivity condition for $\tilde w(x,y)$ may be hard to verify. The opposite way is much easier. In order to compute the classical probability distribution which corresponds to a given quantum state, it is sufficient to compute the Wigner representation of the density matrix $\rho_w(z,p)$ and to check if this is positive.



\begin{thebibliography}{100}
\bibitem{3A}C. Wetterich, arXiv: 0911.1261
\bibitem{CWAA}C. Wetterich, arXiv: 0906.4919, to appear in Annals of Phys.
\bibitem{CW2}C. Wetterich, arXiv: 0810.0985
\bibitem{CWE}C. Wetterich, arXiv:0811.0927, Journal of Phys. {\bf 174} (2009) 012008
\bibitem{3}C. Wetterich, in ``Decoherence and Entropy in Complex Systems'', ed. T. Elze, p. 180, Springer Verlag 2004, arXiv: quant-ph/0212031
\bibitem{DC}H. D. Zeh, Found. Phys. {\bf 1} (1970) 69;\\
E. Joos, H. D. Zeh, Z. Phys. {\bf B59} (1985) 273;\\
E. Joos, H. D. Zeh, C. Kiefer, D. Giulini, J. Kupsch, I.-O.  Stamatescu,
``Decoherence and the appearance of the classical world'', Springer 2003; \\
W. Zurek, Rev. Mod. Phys. {\bf 75} (2003) 715
\bibitem{CW1}C. Wetterich, arXiv: 0809.2671
\bibitem{Bell}J. S. Bell, Physica 1 (1964) 195
\bibitem{BS}J. Clauser, M. Horne, A. Shimony, R. Holt, Phys. Rev. Lett. {\bf 23} (1969) 880;\\
J. Bell, ``Foundations of Quantum Mechanics'', ed. B. d'Espagnat (New York: Academic, 1971) p. 171;\\
J. Clauser, M. Horne, Phys. Rev. {\bf D10} (1974) 526;\\
J. Clauser, A. Shimony, Rep. Prog. Phys. {\bf 41} (1978) 1881
\bibitem{GenStat}C. Wetterich, Nucl. Phys. {\bf B314} (1989) 40; Nucl. Phys. {\bf B397} (1993) 299
\bibitem{EPR}A. Einstein, B. Podolski, N. Rosen, Phys. Rev. {\bf 47} (1935) 777
\bibitem{Ze}D. Bouwmeester, J. W. Pan, K. Mattle, M. Eibl, H. Weinfurter, A. Zeilinger, Nature {\bf 390} (1997) 575
\bibitem{Zo}R. Feynman, Int. J. Theor. Phys. {\bf 21} (1982) 467;\\
D. Deutsch, Proc. R. Soc. London {\bf A400} (1985) 97;\\
J. I. Cirac, P. Zoller, Phys. Rev. Lett. {\bf 74} (1995) 4091
\bibitem{PO}G. Birkhoff, J. von Neumann, The Logic of Quantum mechanics, Vol. 37 (1936);\\
J. von Neumann, ``Mathematical Foundations of Quantum Mechanics'', Princeton University Press (1955);\\
B. Misra, ``Physical Reality and Mathematical Description'', eds. C.~P.~Enz, J.~Mehra (Dordrecht, Reidel, 1974), p.455;\\
A. S. Holevo, ``Probabilistic and Statistical Aspects of Quantum Theory'' (Amsterdam, North Holland, 1982);\\
S. T. Ali, E. Prugovecki, J. Math. Phys. {\bf 18} (1977) 219;\\
M. Singer, W. Stulpe, J. Math. Phys. {\bf 33} (1992) 131;\\
E. Beltrametti, S. Bugajski, J. Phys. A: Math. Gen. {\bf 28} (1995) 3329; Int. J. Theor. Phys. {\bf 34} (1995) 1221;\\
S. Bugajski, Int. J. Theor. Phys. {\bf 35} (1996) 2229;\\
W. Stulpe, P. Busch, J. Math. Phys. {\bf 49} (2008), 3
\bibitem{KS}S. Kochen, E. P. Specker, Journal of Mathematics and Mechanics {\bf 17} (1967), 59;\\
N. D. Mermin, Phys. Rev. Lett. {\bf 65} (1990) 3373;\\
A. Peres, J. Phys. A: Math. Gen. {\bf 24} (1991) L175;\\
N. Straumann, arXiv: 0801.4931
\bibitem{Wig}E. P. Wigner, Phys. Rev. {\bf 40} (1932) 749
\bibitem{Moyal}J. E. Moyal, Proc. Cambridge Philosophical Society {\bf 45} (1949) 99
\bibitem{AW1}G. Aarts, G. F. Bonini, C. Wetterich, Nucl. Phys. {\bf B587} (2000) 403; Phys. Rev. {\bf D63} (2000) 025012
\bibitem{4}C. Wetterich, Phys. Lett. {\bf B399}(1997) 123
\bibitem{CWNEQ}C. Wetterich, Phys. Rev. {\bf E56} (1976) 2687
\end{thebibliography}
\end{document}